\documentclass[a4paper, 12pt]{article}
\usepackage{multicol, multirow, makecell}
\usepackage{geometry}
\usepackage{amsmath, amssymb, amsthm}
\usepackage{bbm}
\usepackage{color}
\usepackage{enumerate} 
\usepackage[inline]{enumitem}
\usepackage[title]{appendix}
\usepackage{refcount}
\usepackage{caption, subcaption}
\usepackage{hyperref}
\usepackage[noabbrev, nameinlink]{cleveref}
\usepackage{calc}
\usepackage{tikz, pgfplots}
\usepgfplotslibrary{fillbetween}

\usepackage[
	backend=biber,
	style=apa,
]{biblatex}
\ExplSyntaxOn
\newcounter{save}
\seq_new:N \g_ephraim_save_seq
\NewDocumentEnvironment{save}{o+b}
{
	\refstepcounter{save}
	\IfValueT{#1}{\label{save@#1}}
	\seq_gput_right:Nn\g_ephraim_save_seq{#2}
}{}
\NewDocumentCommand{\load}{m}
{
	\clist_map_inline:nn {#1}
	{
		\seq_item:Nn\g_ephraim_save_seq
		{
			\getrefnumber{save@##1}
		}
		\par
	}
}
\NewDocumentCommand{\loadall}{}
{
	\seq_use:Nn\g_ephraim_save_seq{\par}
}
\ExplSyntaxOff

\pgfplotsset{compat=1.18}

\newcolumntype{M}[1]{>{\centering\arraybackslash}m{#1}}

\newcommand{\E}{\mathbb{E}}
\newcommand{\D}{\mathrm{d}}
\newcommand{\Eta}{\mathrm{H}}
\newcommand{\switch}[1]{\left\{\begin{aligned}#1\end{aligned}\right.}
\newcommand{\switchplus}[2]{\left\{\begin{array}{#2}#1\end{array}\right.}
\newcommand{\cl}[1]{\mathcal{#1}}
\newcommand{\bs}[1]{\boldsymbol{#1}}
\newcommand{\st}{\text{s.t. }}
\DeclareMathOperator*{\argmax}{\arg\max}
\DeclareMathOperator*{\argmin}{\arg\min}

\numberwithin{equation}{section}
\numberwithin{figure}{section}

\theoremstyle{definition}
\newtheorem{lemma}{Lemma}
\newtheorem{assumption}{Assumption}
\newtheorem{proposition}{Proposition}
\newtheorem{theorem}{Theorem}
\newtheorem{definition}{Definition}
\newtheorem{example}{Example}[section]
\newtheorem{corollary}{Corollary}[theorem]

\newtheorem*{claim*}{Claim}
\newtheorem*{literature*}{Related literature}

\begin{document}

\title{Balancing Selection Efficiency and Societal Costs in Selective Contests\thanks{I would like to thank Federico Echenique, Yingkai Li, Shuo Liu, Matias Nunez, Quitz{\'e} Valenzuela-Stookey, Xi Weng, and Zenan Wu for their helpful comments and suggestions.}}
\author{Penghuan Yan\thanks{Guanghua School of Management, Peking University; \href{mailto:yan.penghuan.2003@gmail.com}{yan.penghuan.2003@gmail.com}}}
\date{}
\maketitle

\begin{abstract}

Selective contests can impair participants' overall welfare in overcompetitive environments, such as school admissions. This paper models the situation as an optimal contest design problem with binary actions, treating effort costs as societal costs incurred to achieve a desired level of selectivity. We provide a characterization for the feasible set of selection efficiency and societal cost in selective contests by establishing their relationship with feasible equilibrium strategies. We find that selection efficiency and contestants' welfare are complementary, i.e. it is almost impossible to improve one without sacrificing the other. We derive the optimal equilibrium outcome given the feasible set and characterize the corresponding optimal contest design. Our analysis demonstrates that it is always optimal for a contest designer who is sufficiently concerned with societal cost to intentionally introduce randomness into the contest. Furthermore, we show that the designer can optimize any linear payoff function by adjusting a single parameter related to the intensity of randomness, without altering the specific structure of the contest. 

{\bf Keywords:} selective contest, selection efficiency, effort cost, competition, school admission

\end{abstract}

\section{Introduction}

Contests are widely employed as mechanisms to identify and select capable individuals. Contest designers incentivize participants to exert costly effort, which is then reflected in their performance, serving as a signal of their underlying ability. However, in highly competitive environments where participants are already well-prepared, the additional effort contributes minimally to enhancing their abilities. In such cases, the effort cost becomes a pure welfare loss for the contestants. When a contestant's performance does not directly translate into revenue for the contest designer, this effort cost can be viewed as a societal cost that the designer must bear to differentiate more capable contestants from those who are less competent.

This trade-off between selection efficiency and societal cost is frequently observed in contexts such as college or graduate school admissions, where students are ranked and selected based on exam scores, GPA, or other comprehensive indices deemed indicative of their ability. In scenarios characterized by intense competition or when the rewards are highly valuable, selective contests are often criticized for driving students to exert excessive effort that is hardly compensated by the benefits of more efficient selection \parencite{chen2022phenomena, li2021involution}.

To investigate how to balance selection efficiency and societal cost, we highlight the use of lotteries in selection processes. A lottery allocates opportunities based on a predetermined probability distribution, disregarding participants' effort, performance, and other potential signals of capability. For instance, in school admissions, admission slots may be randomly distributed among qualified applicants without further assessing their abilities. Compared to contests, lotteries eliminate the need for participants to compete in costly effort, thereby improving their overall welfare. While lotteries are often considered equitable for reducing welfare loss in overcompetitive environments, the most evident drawback of pure lotteries is that they completely surrender selectivity. Due to their inherent randomness, lotteries can fail to identify the most capable candidates, leading to a significant mismatch of resources.

Contests and lotteries represent two extremes of selection methods. Hybrid mechanisms that combine elements of both, such as giving higher-ranked contestants a higher probability of selection without making it one hundred percent, may offer a way to identify capable candidates without excessively compromising participants' welfare. This paper aims to provide an answer to the problem of how to strike a balance between selection efficiency and societal cost in designing selection mechanisms when the principal has a fixed number of positions to allocate.

This problem can be formulated under the framework of contest design with a fixed budget. We consider a contest in which the principal distributes among multiple agents a fixed number of identical, indivisible prizes, such as college admission slots. To simplify the analysis, we assume that agents' actions are binary. Agents can either invest costly effort to achieve a high level of performance or refrain from investing effort, resulting in a low level of performance. The cost of achieving high performance decreases with the agent's inherent ability. In an intensely competitive environment where agents are already well-prepared, we assume that any incurred effort cost does not contribute to improving their abilities and is thus treated as a societal cost.

We allow the principal to condition the allocation rule freely on agents' actions. The principal may choose to allocate prizes to those who perform well before those who perform poorly as in a normal contest, or to disregard agents' performance completely and distribute the prizes equally and randomly with a lottery, or anything in between. The principal's payoff increases when the expected average ability of the agents who are awarded a prize (i.e., those selected) increases, or when the total effort cost incurred during the selection process decreases. Our objective is to determine the optimal selection mechanism that maximizes the principal's payoff.

Our work belongs to the strand of literature on contests that focuses on improving contestants' welfare, in which effort is treated as an undesirable societal cost rather than a source of revenue of the contest designer. Compared to previous work on contestants' welfare, we emphasize the trade-off with contests' ability to allocate prizes to the most capable participants, which we refer to as selection efficiency. Earlier studies have demonstrated that the use of lotteries can enhance contestants' welfare compared to normal contests \parencite{bodoh2018college, krishna2022pareto}. We extend this analysis by examining the comprehensive effect of introducing randomness into contests on both participants' welfare and selection efficiency.

Studies that focus on selection efficiency have examined various modifications on traditional contests to identify capable participants more effectively. Most previous studies in this literature do not care about contestants' effort cost incurred, often proposing contest designs with multiple stages in which participants are motivated to exert effort repeatedly in order to improve selection efficiency \parencite{meyer1991learning,kawamura2014biasing, drugov2017biased, drugov2022selecting}. Our work takes a different approach by treating incurred effort costs as a welfare loss. Rather than maximizing selection efficiency at all costs, we advocate for the intentional introduction of randomness into contest designs to deliberately sacrifice some selection efficiency in order to protect contestants' welfare. This approach is similar to \textcite{li2024screening}, with some distinctions in the principal's payoff structure and constraints on mechanisms.

In this paper, we first characterize the feasible set of selection efficiency and societal cost that a principal can achieve by designing the allocation rule in a multi-agent, multi-prize contest where agents choose between two levels of effort. We do this by demonstrating how agents' equilibrium strategies influence both selection efficiency and societal cost, and by providing a necessary and sufficient condition under which the principal can induce a specific equilibrium strategy via an appropriate contest design.

We then develop a method to determine the optimal feasible equilibrium outcome given the principal's linear preference over societal cost and selection efficiency. We incorporate a technique related to the ``ironing'' method developed by \textcite{myerson1981optimal}. We further provide a characterization for optimal contest designs that induce the desired equilibrium outcome. With two or more prizes to allocate, there is no guarantee that mechanisms which uniquely induce the desired outcome always exist. In the case of a single prize, we present an example of a collection of mechanisms within which a principal with any linear cost-efficiency preference can find a suitable design that induces the optimal outcome as the unique equilibrium.

When agents' actions are binary, they will always play cut-off strategies in equilibrium. Our first major finding is that in a selective contest involving potential randomness, selection efficiency and societal cost are fully reflected in the cut-off of the equilibrium strategy, regardless of the specific contest design and allocation outcome of the prizes. This implies that all feasible cost-efficiency pairs lie on a one-dimensional curve, rather than within a two-dimensional region. Selection efficiency and contestants' welfare are complementary in a sense that it is almost impossible to improve one without reducing the other.

We further show that it is always optimal for the principal to intentionally introduce randomness into the allocation rule as long as the principal places sufficient weight on societal cost. If a contest design involves a one-dimensional parameter that controls the level of randomness involved, the principal can optimize any linear payoff function by adjusting this single parameter without altering the overall contest design. For instance, the principal can deliberately misobserve agents' performance with some probability to mitigate excessive competition. Any optimization objective can be achieved by properly setting the probability of misobservation.

Finally, we provide comparative statics results on how the principal's preferences, the prize-to-participants ratio, and the distribution of agents' ability and effort cost influence the optimal contest design. We find that agents' willingness to invest effort cost under optimal contest design diminishes when the principal places greater emphasis on societal cost, when the prize-participants ratio increases, or when agents' effort costs rise more steeply in percentage as their ability decreases.
 
\begin{literature*}

Our work is mainly related to the literature on optimal contest design with a fixed budget. Pioneered by \textcite{lazear1981rank}, most studies in this literature focus on effort-maximizing contests. \textcite{glazer1988optimal} and \textcite{moldovanu2001optimal} proved that it is optimal to allocate all prizes as a single prize to the top contestant when ability is private information. \textcite{liu2014effort} and \textcite{zhang2024optimal} obtained similar results in settings where the designer is also free to design the contest mechanism. In contrast, when there is complete information of contestants' ability, studies have shown that it is optimal to equally award every contestant but the last one \parencite{barut1998symmetric,letina2023optimal}. Much of this literature treats contestants' effort (performance) as the principal's revenue, akin to a firm manager using tournaments as labor contracts to motivate workers to devote time and effort. \textcite{fang2020turning} found that under complete information settings, increasing competitiveness consistently discourages effort. Their main focus, however, is still on effort maximization.

Little attention has been paid to contestants' welfare in which the effort cost is viewed as a loss. Notable exceptions include \textcite{bodoh2018college}, who examined the welfare cost characterized by an aggregate welfare function in contests such as collection admissions. They found that using a lottery to allocate college seats can result in a higher overall welfare level compared to a contest. \textcite{krishna2022pareto} studied Pareto improvements in students welfare in college admissions within the large contest framework established by \textcite{olszewski2016large}. They extended the results of \textcite{bodoh2018college} by showing that it is better to categorize contestants into several groups according to their performance and then conduct a separate lottery within each group, which they refer to as ``pooling''. Similar to their results, our work also incorporates the idea of introducing randomness into contests to improve social welfare. The major difference between our work and theirs is that we allow the contest designer to have her own evaluation of allocation efficiency, as opposed to their approach in which the allocation outcome affects the social welfare primarily through contestants' evaluation of the prizes they receive, with the designer's payoff being merely an aggregation of contestants' welfare. 

Another important focus of our work is selection efficiency, i.e. whether prizes are allocated to the most capable agents. In many settings, the goal of maximizing effort aligns with the need to select the most capable individuals. Therefore, not much of the literature has  emphasized measuring and improving selection efficiency in contests specifically. Literature that directly regards selection efficiency includes \textcite{clark2001rank}, who studied how selection efficiency can be enhanced in a reward system where the prize depends on the winning effort in a two-agent setting. \textcite{clark2007contingent} further proved that selection efficiency in a contest can be improved when the rewards are endogenized, meaning agents are allowed to choose the prize scheme. Other approaches to improve selection efficiency include reducing the scale of the contest or restricting the quality of participants \parencite{hvide2003risk}. 

Among all studies concerning selection efficiency in contests, the only one we are aware of that analyzed the balance between selection efficiency and agents' utility is \textcite{li2024screening}. They examined the optimal mechanism design for a benevolent social planner allocating limited resources, where unproductive effort serves as a signal. They arrived at a result similar to ours, demonstrating that with a sufficiently large number of agents, the optimal mechanism is always a contest that incorporates randomization. The primary distinction between their work and ours lies in the principal's payoff structure. In their framework, the principal's utility is assumed to depend on the total utility of the agents, which includes the value of the prizes awarded. In other words, the principal incurs a direct penalty for retaining any unallocated prizes. This corresponds to a setting where the principal's resources are dedicated solely to the agents modeled and are considered wasted if not distributed. Consequently, the optimal mechanism in their context naturally involves never withholding prizes. By contrast, in our setting, the principal's payoff is contingent on agents' effort cost. This corresponds to a scenario where the principal's budget is not confined to agent selection, and any unallocated prizes can be used for alternative purposes, resulting in no welfare loss for the principal.

Less similar studies include \textcite{hochtl2011incentives}, who developed a method to trade off selection efficiency and eliciting effort in a setting where they are not perfectly aligned by adjusting the number of heterogeneous participants included. Another study related to our work in terms of its conceptual approach is \textcite{ryvkin2010selection}, who examined how the expected quality of prize winners changes with varying noise levels. In our work, the principal can select allocation rules that deliberately introduce such noise to balance selection efficiency and societal cost.

A strand of literature on selection efficiency that is less relevant to our work focuses on introducing bias or headstarts into contests to increase the likelihood of awarding the prize to the most capable agent \parencite{meyer1991learning,kawamura2014biasing, drugov2022selecting}. \textcite{drugov2017biased} found similar results in settings with noise. Such improvements typically require the contest to have multiple stages, which can significantly harm the contestants due to the repeatedly incurred effort cost. In our setting, however, the principal has to balance selection efficiency and societal cost within a single-stage contest.

There is also a substantial body of empirical research examining the impact of contests and lotteries in real-world selection processes, such as school admissions. Studies have shown that admission contests, which rank students by their test scores and GPA, can lead to academic anxiety in highly competitive environments \parencite{li2021involution, chen2022phenomena}. Additionally, under such contests, students may avoid taking challenging courses that do not contribute to a high GPA but are beneficial for their long-term development, ultimately harming their long-term welfare \parencite{zhang2023college}. While admission lotteries are generally believed to improve student welfare, \textcite{schripsema2014selection} pointed out that students selected via lottery tend to perform worse than those selected through contests, as they may not fully utilize educational opportunities to their maximum potential.

\end{literature*}

The rest of the paper is organized as follows. In \Cref{sec:model}, we present a model of selective contests with incomplete information and binary actions, and develop our measures for selection efficiency and societal cost in such contests. \Cref{sec:cutoff} establishes the relationship between the symmetric Bayesian Nash equilibrium and its associated selection efficiency and societal cost. In \Cref{sec:feasible}, we characterize the feasible set of equilibrium strategy profiles and mechanisms sufficient to induce this feasible set. In \Cref{sec:optimal}, we provide an approach to solving for the optimal selective contest given the feasible set. \Cref{sec:comparative} presents some comparative statics results with an example. \Cref{sec:conclusion} concludes. 

\section{Model}
\label{sec:model}
\subsection{The Contest}

We consider a contest in which the principal allocates $m$ identical, indivisible prizes (positions) to $n$ risk neutral agents ($0<m<n$). The agents that receive the prizes are considered ``selected''. Denote the set of agent by $N=\{1,2,\dots,n\}$. Each agent is characterized by her type (ability) $\theta\in[0,1]$, private information known solely to the agent herself. The agents' types are drawn randomly from the same distribution with c.d.f. $F$ independently. 

In the contest, each agent chooses to exert either high effort ($H$) or low effort ($L$). The action space of an agent is denoted by $A=\{H,L\}$. Opting for high effort incurs a cost of $c(\theta)$ for a type-$\theta$ agent. We assume that $c(\theta) \geq 0$ for every type of agent and that $c(\theta)$ is strictly increasing in $\theta$, indicating that agents with lower type values possess greater ability. Choosing low effort carries no cost. For simplicity, we do not differentiate between effort and performance here. The principal has perfect visibility into each agent's effort or, equivalently, the observed performance levels of the agents are perfectly correlated with their efforts.
 
The principal distributes prizes to the agents by designing an allocation rule. The allocation rule operates on the action profile of the agents and generates a lottery, the outcome of which is an $n$-dimensional vector, where the $i$-th component denotes the number of prizes allocated to agent $i$. Because agents are risk neutral, it is tantamount to considering the interim allocation before the lottery is realized. Consequently, we express the mechanism as $q:A^n\to \mathbb{R}_+^n$, where $q_i(a)$ signifies the expected amount of prize that agent $i$ receives when the action profile is $a$.

To capture the nature of selective contests, we impose three assumptions on the mechanism available to the principal.

\begin{assumption}
	\label{ass_useup}
	For all $a\in A^n$, $\sum_{i\in N} q_i(a) = m.$
\end{assumption}

\Cref{ass_useup} requires the principal to never keep prizes. Most of the real-world selective contests are more or lees subject to this constraint. For example, promotion opportunities arise when vacant positions must be filled to maintain an organization's normal operations. In college entrance exams, admission quotas are typically set before test scores are revealed. Although the principal may have limited flexibility in prize allocation, significant deviation from the quota is not feasible. Therefore, for simplicity, we assume that the principal strictly adheres to the quota and consistently allocates $m$ prizes.

Note that $q$ is implemented by a lottery over $\{0,1\}^n$, and ensuring that exactly $m$ prizes are always allocated ex post requires that every possible outcome of the lottery sums up to $m$ prizes. It can be shown that this requirement is equivalent to \Cref{ass_useup}; that is, every $q(a)$ that satisfies \Cref{ass_useup} can be supported by some lottery over $\{v\in\{0,1\}^n:\sum_{i\in N} v_i = m\}$.

\begin{assumption}
	\label{ass_sym}
	For every permutation $\sigma:N\to N$, $q_{\sigma(i)}(a_{\sigma(1)}, a_{\sigma(2)}, \dots, a_{\sigma(n)}) = q_i(a)$, $\forall a\in A^n$, $i\in N$.
\end{assumption} 

Under \Cref{ass_sym}, the principal is restricted to using symmetric allocation rules and breaking ties evenly. In real-world scenarios, the principal often receives additional signals beyond effort and performance, which can inform tie-breaking decisions. However, in our simplified model, the principal only observes the action profile. Thus, this assumption is made purely for convenience.

\begin{assumption}
	\label{ass_bound}
	For all $i\in N$, $a\in A^n$, $0\leq q_i(a)\leq1$.
\end{assumption}

\Cref{ass_bound} encapsulates a key characteristic of selective contests, such as employee promotions and college entrance exams. Each agent can only receive one of two possible ex-post outcomes: selected or not selected. The principal can adjust the expected value of the prize to any amount between $0$ and $1$ through randomization, though the value cannot exceed $1$.

If \Cref{ass_bound} does not hold, we can normalize the total value of prizes (positions) to $1$ by scaling the cost function $c$. Consequently, this situation is equivalent to the case where $m=1$.

In real-world selective contests, such as college admissions, the amount of scholarship is often relatively insignificant compared to the value of the college offer itself. Students are unlikely to forgo college enrollment over the relatively small amount of a typical scholarship. Therefore, we exclude monetary transfers from the model.

Agent $i$'s payoff when the action profile is $a$ is given by
\begin{equation*}
	q_i(a)-c(\theta_i)\cdot \mathbbm{1}[a_i=H].
\end{equation*}

\subsection{Principal's Problem}

The goal of a selective contest is to allocate prizes to those with greater ability, which corresponds to agents with lower types in this model. To achieve this, the principal must induce observable effort choices from the agents as a signal of their types. We assume that the agent's effort in preparing for the contest does not significantly enhance their inherent ability. Thus, the effort cost is considered a societal expense necessary for selecting the most suitable candidates, representing a pure welfare loss. This distinguishes our work from previous contest literature, which typically views agents' effort and performance as benefiting the principal, similar to a production setting where a manager incentivizes workers to maximize productivity.

The principal's ultimate goal is to maximize selection efficiency while minimizing societal cost. We assume that the principal's payoff is linear in selection efficiency, denoted by $\eta$, and societal cost, denoted by $C$. The principal's problem is therefore formulated as:
\begin{equation*}
\max_{q}\ \eta - \lambda C,
\end{equation*}
where $\lambda > 0$ is a predetermined constant representing the principal's aversion to societal cost.\footnote{In real-world selective contests, contestants' efforts can be productive in improving their abilities. Within the framework of our model, $\lambda \leq 0$ represents the case where the contestants' efforts are sufficiently beneficial that the principal perceives their impact as a net gain. As discussed in the Literature Review, the objectives of inducing effort and improving selection efficiency are often aligned. When $\lambda \leq 0$, the problem becomes less interesting, as it is nearly always optimal for the principal to implement a standard contest, which provides strong incentives for capable contestants to exert effort, thereby generating a sharp signal that allows the principal to identify them. Consequently, for the remainder of this paper, we focus on the case where $\lambda > 0$. It is important to note that our subsequent approach to deriving the optimal contest design does not depend on this assumption. The method presented can still be applied to determine the optimal mechanism when $\lambda \leq 0$, although the resulting optimal outcome will most likely be the standard contest.}

We characterize the selection efficiency with an affine transformation of the expected average type of selected agents.\footnote{To justify that an affine transformation of selected agents' average type is a reasonable measure for selection efficiency, we start by assuming that a selected agent with type $\theta$ provides a benefit $b(\theta)$ to the principal. Since a lower type indicates a more capable agent, $b$ should be strictly decreasing in $\theta$. We further assume that the benefit from each selected agent is additive. This is analogous to measuring market (Pareto) efficiency using producer and consumer surplus, which is additive across individuals. We relabel the agents' types and scale $b$ so that $b(\theta) = -\theta$. Consequently, the total benefit of selecting all agents in $M \subset N$ is given by $-\sum_{i\in M} \theta_i$. This is equal to $-m\E[\text{selected agents' average type}]$, an affine transformation of selected agents' average type.} A low expected average type implies a high overall ability of selected agents, which then indicates a high selection efficiency. We normalize the efficiency of a completely random mechanism to $0$ and the efficiency of a mechanism that guarantees to allocate all prizes to agents with type $0$, which represents the highest ability, to $1$. This yields the following definition of selection efficiency.

\begin{definition}[Selection efficiency]
	The {\it selection efficiency} of the contest, denoted by $\eta$, is defined as
	\begin{equation*}
		\eta:=1-\frac{\E[\text{selected agents' average type}]}{\E[\text{agents' average type}]}=1-\E\left[\frac{\sum_{i\in N}\theta_i\cdot q_i(a(\theta))}{m\mu}\right],
	\end{equation*}
	where $\mu=\E_{\theta\sim F}[\theta]$.
\end{definition}

The societal cost is represented by the expected total effort cost of the agents. To control for the effect of the contest scale $n$ on the total cost measure, we characterize the societal cost by the expected average effort cost induced by the contest.

\begin{definition}[Societal cost]
	The {\it societal cost} of the contest, denoted by $C$, is defined as
	\begin{equation*}
		C:=\E\left[\frac{\sum_{i\in N} c(\theta_i)\cdot\mathbbm{1}[a_i=H]}{n}\right]=\E_{\theta_i\sim F}[c(\theta_i)\cdot\mathbbm{1}[a_i=H]].
	\end{equation*}
\end{definition}

\section{The Cut-off Property}
\label{sec:cutoff}

Given that the action set is binary, allowing agents to report their type reveals more information than what the principal can infer from the agents' action profile. Therefore, rather than optimizing over all possible mechanisms under incentive compatibility constraints, we begin by characterizing the Bayesian Nash equilibrium of the game. For simplicity, we assume that if an agent is indifferent between exerting high or low effort, she will always choose to exert low effort.

\begin{lemma}
	\label{cutoff}
	The only existing symmetric equilibria are cut-off equilibria, in which the agents play the same cut-off strategy that takes the following form:
	\begin{equation*}
		a_i = \switch{
			&H, &\theta_i & < s \\
			&L, &\theta_i & \geq s \\
		},
	\end{equation*}
	where $s\in[0,1]$ is the cut-off.
\end{lemma}

\begin{proof}
	Suppose the strategy profile played by all agents but agent $i$ is $s_{-i}$. Denote the interim allocation agent $i$ faces when her action is $a_i$ and other agents are playing strategy $s_{-i}$ by $Q_i(a_i, s_{-i})$. Then, agent $i$ with type $\theta_i$ will exert high effort if and only if
	\begin{equation*}
		Q_i(H, s_{-i}) - Q_i(L, s_{-i}) > c(\theta_i).
	\end{equation*}
	Therefore, the best response of agent $i$ is always a cut-off strategy that can be expressed as
	\begin{equation}
		\label{BR}
		\mathrm{BR}_i(s_{-i}) = \switch{
			&H, &\theta_i & < c_{-1}(Q_i(H, s_{-i}) - Q_i(L, s_{-i}))\\
			&L, &\theta_i & \geq c_{-1}(Q_i(H, s_{-i}) - Q_i(L, s_{-i}))\\
		},
	\end{equation}
	where $c_{-1}$ is defined as
	\begin{equation*}
		c_{-1}(x) := \switch{
			&0, &x &<c(0) \\
			&c^{-1}(x), &x&\in[c(0),c(1)] \\
			&1, &x &>c(1) \\
		}.
	\end{equation*}
	Hence, the only existing symmetric equilibria are cut-off equilibria.
\end{proof}

The cut-off property arises due to the monotonicity of the cost function $c$. With a slight abuse of notation, we also use $s$ to denote the strategy itself. \Cref{cutoff} indicates that every symmetric equilibrium can be characterized by a real number $s \in [0,1]$, representing the strategy of the agents in equilibrium. We refer to this as the {\it equilibrium cut-off}.

According to \eqref{BR}, $s$ is a valid equilibrium cut-off if and only if
\begin{equation}
	\label{eq_cutoff_cond_tmp}
	c_{-1}(Q_i(H,\bs{s}) - Q_i(L,\bs{s})) = s,
\end{equation}
where $\bs s = (\overbrace{s,s,\dots,s}^{n-1})$. 
Denote by $\phi(s,q)$ the difference in payoff for an agent with type $s$ between exerting low effort and high effort, given that everyone else is playing a strategy with a cut-off also equal to $s$ under mechanism $q$. Formally,
\begin{equation}
	\label{def_phi}
	\phi(s,q):= Q_i(H,\bs{s}) - c(s) - Q_i(L,\bs{s}).
\end{equation}
Then, the condition given by \eqref{eq_cutoff_cond_tmp} can be expressed in terms of $\phi$ as
\begin{equation}
	\label{eq_phi}
	\begin{aligned}
		\phi(s,q) &= 0, &\text{if }s&\in(0,1),\\
		\phi(s,q) &\leq 0, &\text{if }s&=0,\text{ and}\\
		\phi(s,q) &\geq 0, &\text{if }s&=1.\\
	\end{aligned}
\end{equation}

$\phi$ offers a straightforward characterization of the equilibrium cut-offs. Specifically, it captures the agents' incentive to deviate when everyone is playing the same cut-off strategy. If $\phi > 0$, agents are incentivized to deviate to a higher cut-off; if $\phi < 0$, they are incentivized to deviate to a lower cut-off.

\begin{proposition}
	\label{prop_exist}
	If $c$ and $F$ are continuous on $[0,1]$, then for any possible mechanism $q$, there exists at least one symmetric equilibrium.
\end{proposition}

\begin{proof}
	See \Cref{app:prop_exist}.
\end{proof}

\begin{save}[prop_exist_pf]
\begin{proof}
	With the assumption that $c$ and $F$ are continuous, $\phi$ is continuous in $s$.

	Consider $\phi(0,q)$ and $\phi(1,q)$. If $\phi(0,q)\cdot \phi(1,q)<0$, then by intermediate value theorem of continuous functions, there exists at least one $s\in(0,1)$ that satisfies $\phi(s,q)=0$. According to \eqref{eq_phi}, this indicates that $s$ is an equilibrium cut-off under $q$.

	If $\phi(0,q) \cdot \phi(1,q) \geq 0$, then at least one of $\phi(0,q)\leq0$ and $\phi(1,q)\geq0$ must hold. According to \eqref{eq_phi}, the former implies that $s=0$ is an equilibrium cut-off, and the latter implies that $s=1$ is an equilibrium cut-off under $q$.

	Therefore, given that $c$ and $F$ are continuous on $[0,1]$, there exists at least one symmetric equilibrium under any allocation rule $q$.
\end{proof}
\end{save}

The existence of symmetric equilibrium allows us to evaluate each mechanism based on the equilibrium cut-offs it can induce. This leads to our first major result, which explicitly characterizes selection efficiency and societal cost.

\begin{theorem}
	\label{thm_cutoff}
	In a symmetric equilibrium, societal cost $C$ and selection efficiency $\eta$ are determined only by the equilibrium cut-off, but not the detailed structure of mechanism $q$, given the cost structure $c$ and type distribution $F$.
\end{theorem}

\begin{proof}[Proof sketch]
	{\bf Societal cost.} In an equilibrium with cut-off $s$, an agent exerts high effort if and only if her type is below $s$. Therefore, by definition, the societal cost $C$ can be expressed as
	\begin{equation*}
		C = \E_{\theta_i\sim F}[c(\theta_i)\cdot \mathbbm{1}[\theta_i<s]],
	\end{equation*}
	which is fully determined by $s$, given $c$ and $F$.

	{\bf Selection efficiency.} We categorize the agents into two groups: those exerting high effort and those exerting low effort. This division is determined solely by the cut-off $s$. By definition, the selection efficiency $\eta$ is an affine transformation of the expected type of the selected agents, which can be expressed using the expected type of agents in the two categories and the probability of a selected agent belonging to each category.
	
	The expected type within each category is determined by $s$, given $c$ and $F$. The probability of a selected agent being in each category can be expressed in terms of $Q_i(H,\bs s)$ and $Q_i(L,\bs s)$ via Bayes' theorem. The weighted average of these probabilities, with the probability of being in each category as the weight, equals the ex-ante probability of being selected, $m/n$. Together with \eqref{eq_phi}, this forms a system of linear equations in two unknowns that uniquely determines $Q_i(H,\bs s)$ and $Q_i(L,\bs s)$ in terms of $s$, given $c$ and $F$. This, in turn, determines $\eta$, provided that $s \in (0,1)$. When $s = 0$ or $1$, all agents behave identically, resulting in $\eta = 0$.
	
	It can be shown that given $c$ and $F$, societal cost $C$ and selection efficiency $\eta$ can be expressed in terms of the equilibrium cut-off $s$ as 
	\begin{equation}
		\label{C_exp_cutoff}
		C = F(s)\E_{\theta\sim F}[c(\theta)|\theta<s] = \int_0^s c(\theta)\D F(\theta), 
	\end{equation}
	and
	\begin{equation}
		\label{eta_exp_cutoff}
		\eta = \frac n{m\mu} c(s)F(s)\left(\mu-\E_{\theta\sim F}[\theta|\theta<s]\right) = \frac {nc(s)}{m\mu}\int_0^s(\mu-\theta)\D F(\theta). \\
	\end{equation}
	See \Cref{app:thm_cutoff} for a detailed proof.
\end{proof}

\begin{save}[thm_cutoff_pf]
\begin{proof}
	{\bf Societal cost.} Let the equilibrium cut-off be $s$. The societal cost $C$ satisfies
	\begin{equation*}
		\begin{aligned}
			C&=\E_{\theta_i\sim F}[c(\theta_i)\cdot\mathbbm{1}[a_i=H]]\\
			&=\E_{\theta_i\sim F}[c(\theta_i)\cdot\mathbbm{1}[\theta_i<s]]\\
			&=F(s)\E_{\theta\sim F}[c(\theta)|\theta<s] \\
			&=\int_0^s c(\theta)\D F(\theta),
		\end{aligned}
	\end{equation*}
	which has $s$ as its only variable. 
	
	{\bf Selection efficiency} As for selection efficiency, \eqref{eq_phi} suggests that for every $s\in(0,1)$,
	\begin{equation}
		\label{QQ_1}
		Q_i(H,\bs s) - Q_i(L,\bs s) - c(s) = 0
	\end{equation}
	holds for all $i\in N$. 
	
	In an equilibrium, for agent $i$, her ex-ante probability of being selected (before she knows her own type) is $m/n$ because agents are symmetric. She will then either be assigned a type $\theta_i<s$ and exert high effort with probability $F(s)$, or be assigned a type $\theta_i\geq s$ and exert low effort with probability $1-F(s)$. Thus, 
	\begin{equation}
		\label{QQ_2}
		\frac mn = F(s)\cdot Q_i(H,\bs s)+(1-F(s))\cdot Q_i(L,\bs s).
	\end{equation}

	When $s\in(0,1)$, \eqref{QQ_1} and \eqref{QQ_2} together pin $Q_i(H,\bs s)$ and $Q_i(L,\bs s)$ down to
	\begin{equation*}
		\switch{
			Q_i(H,\bs s)&=\frac mn + (1-F(s))c(s)\\
			Q_i(L,\bs s)&=\frac mn - F(s)c(s)\\
		}.
	\end{equation*}

	Denote the event ``agent $i$ is selected'' by $W_i$. $Q_i(H,\bs s)$ is equal to the probability of agent $i$ being selected conditioned on that she exerted high effort, and can thus be expressed as
	\begin{equation*}
		Q_i(H,\bs s)=\Pr(W_i|a_i=H).
	\end{equation*}
	By Bayes' theorem, given the equilibrium cut-off,
	\begin{equation*}
		\begin{aligned}
			\Pr(a_i=H|W_i)&=\frac{\Pr(W_i|a_i=H)\cdot \Pr(a_i=H)}{\Pr(W_i)}\\
			&=\frac{Q_i(H,\bs s)\cdot F(s)}{m/n}.
		\end{aligned}
	\end{equation*}
	Similarly, 
	\begin{equation*}
		\Pr(a_i=L|W_i)=\frac{Q_i(L,\bs s)\cdot(1-F(s))}{m/n}.
	\end{equation*}
	Hence, the expected average type of select agents $\E[\theta_i|W_i]$ is given by
	\begin{equation*}
		\label{Etype}
		\begin{aligned}
			\E[\theta_i|W_i] &= \Pr(a_i=H|W_i)\cdot\E[\theta_i|a_i=H]+\Pr(a_i=L|W_i)\cdot\E[\theta_i|a_i=L]\\
			&= \frac{Q_i(H,\bs s)F(s)}{m/n}\int_0^s\theta\D\frac{F(\theta)}{F(s)} + \frac{Q_i(L,\bs s)(1-F(s))}{m/n}\int_s^1\theta\D\frac{F(\theta)}{1-F(s)} \\
			&=\left(1+\frac nm (1-F(s))c(s)\right)\int_0^s\theta\D F(\theta)+\left(1-\frac nm F(s)c(s)\right)\int_s^1\theta\D F(\theta)\\
			&=\left(1-\frac nm F(s)c(s)\right)\mu+\frac nm c(s)\int_0^s\theta\D F(\theta)\\
			&= \mu-\frac nm c(s)\int_0^s(\mu-\theta)\D F(\theta).
		\end{aligned}
	\end{equation*}
	
	By definition, selection efficiency $\eta$ is an affine transformation of the expected average type of select agents, that is,
	\begin{equation}
		\label{eta_exp_cutoff_proved}
		\begin{aligned}
			\eta &= 1-\frac{\E[\text{selected agents' average type}]}{\E[\text{agents' average type}]}\\
			&= 1-\frac{\E[\theta_i|W_i]}{\mu} \\
			&= \frac{nc(s)}{m\mu}\int_0^s(\mu-\theta)\D F(\theta),
		\end{aligned}
	\end{equation}
	which has $s$ as its only variable. 
	
	When $s=0$ or $s=1$, all agents will choose the same action with probability $1$, and thus the selection efficiency is $0$. Hence, \eqref{eta_exp_cutoff_proved} holds for all $s\in[0,1]$.
\end{proof}
\end{save}

\Cref{thm_cutoff} shows that all the information the principal cares about regarding the mechanism is encapsulated in the equilibrium cut-off it induces. Once the cut-off is determined, the societal cost and selection efficiency of the mechanism can be derived without needing additional assumptions about the mechanism's specifics. This implies that the cost-efficiency pair $(C,\eta)$ forms a curve, with the cut-off as the parameter in its parametric equation, rather than a two-dimensional region. In other words, the principal cannot ``do worse.'' The principal selects an equilibrium cut-off and, accordingly, a societal cost as an investment, and then obtains the corresponding selection efficiency as the return.

The cut-off property simplifies the problem of finding optimal contests into two separated steps:

\begin{enumerate}
	\item Find the feasible set of equilibrium cut-offs, denoted by $\cl S$, in which every cut-off can be induced by some mechanism $q$.
	\item Determine the maximizer $s^*$ that solves 
	\begin{equation*}
		\max_s\ \eta-\lambda C\quad\st s\in\cl S,
	\end{equation*}
	and find the corresponding mechanism $q^*$.
\end{enumerate}

This two-step approach is more practical than directly maximizing the principal's utility over a set of mechanisms because the principal cannot ask the agents to report their types. As a result, the first step in optimizing over mechanisms involves finding the Bayesian Nash equilibria for each mechanism, which naturally leads us back to this two-step method.

\section{Feasible Set of Equilibrium Cut-offs}
\label{sec:feasible}

To identify the all feasible equilibrium cut-offs that can be realized under some mechanism, we first provide an explicit characterization of the mechanisms that satisfy the assumptions outlined in \Cref{sec:model}.

\begin{lemma}
	\label{mech}
	Every allocation rule $q$ that satisfies \Cref{ass_useup} and \labelcref{ass_sym} can be characterized by an $(n-1)$-dimensional vector $\bs v \in [0,m]^{n-1}$, where the $k$-th component $v_k$ represents the total expected value of the prize allocated to all agents who exert high effort, given that $k$ agents are exerting high effort. Specifically, for all $i \in N$, the allocation rule is given by:
	\begin{equation*}
		q_i(a) = \switch{
			&\makebox[\widthof{$\dfrac{m-v_k}{n-k},$}][c]{$\dfrac{v_k}{k},$} & a_i&=H \\
			&\frac{m-v_k}{n-k}, & a_i&=L\\
		},
	\end{equation*}
	where $k = |{i : a_i = H}|$ is the number of agents exerting high effort. Additionally, we define $v_0 = 0$ and $v_n = m$.
\end{lemma}

\begin{proof}
	Under the symmetry assumption, the principal must treat agents who take the same action equally. The only information the principal can obtain from the agents' action profiles, and on which the allocation rule can be conditioned, is the number of agents exerting high (or low) effort. Consequently, every mechanism is equivalent to a ``direct'' mechanism that operates as follows:
	\begin{enumerate}
		\item The principal observes the number of agents exerting high effort, denoted by $k$, and chooses a number $0 \leq v_k \leq m$ as the total amount of prizes to allocate among those agents. According to the no-keeping assumption, the remaining $m - v_k$ prizes are allocated to those who exert low effort.
		\item The prizes within each group of agents who chose the same effort level are then distributed evenly, possibly through a fair lottery.
	\end{enumerate}
	As a result, agents who exert high effort receive $v_k/k$ prizes each, while those who exert low effort receive $(m - v_k)/(n - k)$ prizes each.
	
	Furthermore, when no agents exert high effort, the principal must allocate all prizes to the agents exerting low effort; conversely, when all agents exert high effort, the principal must allocate all prizes to them. This implies that $v_0 = 0$ and $v_n = m$ always hold, so we do not need to explicitly include these in our characterization of mechanisms. Therefore, a mechanism can be fully characterized by an $(n-1)$-dimensional vector $\bs v = (v_1,v_2,\dots,v_{n-1})\in[0,m]^{n-1}$.
\end{proof}

We refer to the corresponding $(n-1)$-dimensional vector of a mechanism as its {\it allocation vector}. \Cref{mech} establishes an equivalence between mechanisms and their allocation vectors. Hence, we do not differentiate between the mechanism $q$ itself and its allocation vector $\bs v$ in notation.

\begin{example}
	A mechanism is termed a ``standard contest'' if the principal always selects as many agents as possible who exert high effort before selecting those who exert low effort. Conversely, a mechanism is termed a ``reversed contest'' if the principal always selects as many agents as possible who exert {\it low} effort before selecting those who exert {\it high} effort. The allocation vectors corresponding to a standard contest and a reversed contest, denoted by $\bar{\bs v}$ and $\underline{\bs v}$ respectively, are given by
	\begin{equation*}
		\begin{aligned}
			\bar{\bs v} &= (1,2,\dots,m-1,\underbrace{m,m,\dots,m}_{n-m}), \\
			\underline{\bs v} &= (\underbrace{0,0,\dots,0}_{n-m},1,2,\dots,m-1).
		\end{aligned}
	\end{equation*}
\end{example}

The standard and reversed contests establish the upper and lower bounds for mechanisms that satisfy \Cref{ass_bound} in addition to \Cref{mech}, which stipulates that each agent may receive no more than 1 unit of the prize. Every mechanism that meets all three assumptions outlined in \Cref{sec:model} belongs to the set $\{\bs v: \underline{\bs v}\leq\bs v\leq\bar{\bs v}\}$.

\begin{example}
	A mechanism is termed a ``completely random allocation'' if all agents are selected with equal probability regardless of their effort level. Its allocation vector $\bs v_r$ is given by
	\begin{equation*}
		\bs v_r = \left(\frac mn, \frac {2m}n, \dots, \frac {(n-1)m}n\right).
	\end{equation*}
\end{example}

With the characterization of mechanisms, we provide a sufficient and necessary condition for the feasibility of equilibrium cut-offs.

\begin{theorem}
	\label{thm_feasible}
	Let $\cl S(\bs v_{\min}, \bs v_{\max})$ denote the set of all feasible equilibrium cut-offs that can be induce by using an allocation vector $\bs v$ such that $\bs v_{\min}\leq\bs v\leq\bs v_{\max}$. Then, $s\in \cl S(\bs v_{\min}, \bs v_{\min})$ if and only if
	\begin{equation*}
		\begin{aligned}
			\phi(s,\bs v_{\min})&\leq 0, & \text{if } s&\in[0,1),\text{ and}\\
			\phi(s,\bs v_{\max})&\geq 0, & \text{if } s&\in(0,1],
		\end{aligned}
	\end{equation*}
	where $\phi$ is defined as in \eqref{def_phi}.
\end{theorem}

\begin{proof}
	See \Cref{app:thm_feasible_pf}.
\end{proof}

\begin{save}[thm_feasible_pf]
\begin{proof}
	We start the proof by deriving the explicit expression of $\phi(s,\bs v)$ using $v_k$, $k=1,2,\dots,n-1$. Suppose every agent but agent $i$ is playing the same strategy with cut-off $s$. The probability of exactly $k$ agents (not including agent $i$) exerting high effort is given by
	\begin{equation*}
		\Pr(\text{exactly $k$ agents exert high effort})=(F(s))^k(1-F(s))^{n-1-k}\binom{n-1}{k}.
	\end{equation*}
	If agent $i$ also exerts high effort when there are $k$ other agents exerting high effort, her payoff will be equal to $v_{k+1}/(k+1)$. If she chooses to exert low effort instead, her payoff will be $(m-v_k)/(n-k)$. Hence, the interim allocation rule under $\bs v$ is given by
	\begin{equation*}
		Q_i(H,\bs s) = \sum_{k=0}^{n-1} (F(s))^k(1-F(s))^{n-1-k}\binom{n-1}{k} \frac{v_{k+1}}{k+1},
	\end{equation*}
	and
	\begin{equation*}
		Q_i(L,\bs s) = \sum_{k=0}^{n-1} (F(s))^k(1-F(s))^{n-1-k}\binom{n-1}{k} \frac{m-v_k}{n-k}.
	\end{equation*}
	
	Recall the definition of $\phi$ given in \eqref{def_phi},
	\begin{equation*}
		\phi(s,\bs v)=Q_i(H,\bs s)-c(s)-Q_i(L,\bs s).\tag{\ref{def_phi}}
	\end{equation*}
	With the explicit expression of $Q_i$, we can write $\phi$ as
	\begingroup\allowdisplaybreaks
	\begin{align*}
		\phi(s,\bs v) &= \sum_{k=0}^{n-1} (F(s))^k(1-F(s))^{n-1-k}\binom{n-1}{k} \frac{v_{k+1}}{k+1} - c(s) \\&\qquad- \sum_{k=0}^{n-1} (F(s))^k(1-F(s))^{n-1-k}\binom{n-1}{k} \frac{m-v_k}{n-k}\\
		&= \sum_{k=1}^{n} (F(s))^{k-1}(1-F(s))^{n-k}\binom{n-1}{k-1} \frac{v_{k}}{k} - c(s) \\&\qquad- \sum_{k=0}^{n-1} (F(s))^k(1-F(s))^{n-1-k}\binom{n-1}{k} \frac{m-v_k}{n-k}\\
		&= \frac 1n\sum_{k=1}^{n} (F(s))^{k-1}(1-F(s))^{n-k}\binom{n}{k} v_k - c(s)\\&\qquad- \frac 1n\sum_{k=0}^{n-1} (F(s))^k(1-F(s))^{n-1-k}\binom{n}{k}(m-v_k) \\
		&= \frac 1n\sum_{k=1}^{n-1} (F(s))^{k-1}(1-F(s))^{n-k}\binom{n}{k} v_k + \frac mn (F(s))^{n-1} - c(s)\\&\qquad- \frac 1n\sum_{k=1}^{n-1} (F(s))^k(1-F(s))^{n-1-k}\binom{n}{k}(m-v_k) - \frac mn (1-F(s))^{n-1}\\
		&= \frac 1n\sum_{k=1}^{n-1} (F(s))^{k-1}(1-F(s))^{n-1-k}\binom{n}{k} v_k - c(s)\\&\qquad- \frac mn\sum_{k=0}^{n-1} (F(s))^k(1-F(s))^{n-1-k}\binom{n}{k} + \frac mn (F(s))^{n-1}\\
		&= \frac 1n\sum_{k=1}^{n-1} (F(s))^{k-1}(1-F(s))^{n-1-k}\binom{n}{k} v_k - c(s)\\&\qquad- \frac m{n(1-F(s))}\sum_{k=0}^{n-1} (F(s))^k(1-F(s))^{n-k}\binom{n}{k} + \frac mn (F(s))^{n-1}\\
		&= \frac 1n\sum_{k=1}^{n-1} (F(s))^{k-1}(1-F(s))^{n-1-k}\binom{n}{k} v_k - c(s)\\&\qquad- \frac m{n(1-F(s))}(1-(F(s))^n) + \frac mn (F(s))^{n-1}\\
		&= \frac 1n\sum_{k=1}^{n-1} (F(s))^{k-1}(1-F(s))^{n-1-k}\binom{n}{k} v_k - c(s) - \frac mn\cdot\frac{1-(F(s))^{n-1}}{1-F(s)},\\
	\end{align*}
	\endgroup
	or alternatively, 
	\begin{equation}
		\label{exp_phi}
		\phi(s,\bs v) = \frac 1n\sum_{k=1}^{n-1} (F(s))^{k-1}(1-F(s))^{n-1-k}\binom{n}{k} v_k - c(s) - \frac mn\sum_{k=0}^{n-2}(F(s))^k.
	\end{equation}

	We know from \eqref{exp_phi} that $\phi(s,\bs v)$ is continuous and increasing in $v_k$ for all $k=1,2,\dots,n-1$. This implies that $\phi(s,\bs v_{\min})\leq\phi(s,\bs v)\leq\phi(s,\bs v_{\max})$ for all $\bs v_{\min}\leq\bs v\leq\bs v_{\max}$. For a given $s\in(0,1)$, by \eqref{eq_phi}, $s$ is an equilibrium cut-off under $\bs v$ is equivalent to $\phi(s,\bs v)=0$. Therefore by intermediate value theorem, there exists $\bs v$ such that $s$ is an equilibrium cut-off under $\bs v$ if and only if $\phi(s,\bs v_{\min})\leq0\leq\phi(s,\bs v_{\max})$. Moreover, $0$ is an equilibrium cut-off if and only if $\phi(0,\bs v)\leq0$. The existence of such $\bs v$ implies $\phi(0,\bs v_{\min})\leq\phi(0,\bs v)\leq0$, which in turn indicates that $0$ can be induced by $\bs v_{\min}$. Similarly, $1$ is an equilibrium cut-off under some $\bs v$ such that $\bs v_{\min}\leq\bs v\leq\bs v_{\max}$ if and only if $\phi(1,\bs v_{\max})\geq0$.

	Combining all three cases we obtain a necessary and sufficient condition for $s$ to be in $\cl S(\bs v_{\min},\bs v_{\max})$.
\end{proof}
\end{save}

Intuitively, an increase in any component of the allocation vector indicates that the mechanism favors agents who exert high effort more. When such an increase occurs, the probability that an agent exerting high effort is selected becomes larger, at least in some cases. This increase is reflected in the $\phi$ function, which captures agents' motivation to deviate to a higher cut-off. As a result, they are more likely to exert high effort when everyone is playing the same cut-off strategy. If $\bs v_1 < \bs v_2$, agents will always have more incentive to exert high effort under $\bs v_2$ than under $\bs v_1$, so $\phi(s,\bs v_2) \geq \phi(s,\bs v_1)$ always holds.

In the context of \Cref{thm_feasible}, $\bs v_{\min}$ represents the mechanism that favors high-effort agents the least, while $\bs v_{\max}$ represents the mechanism that favors them the most. When $s \in (0,1)$, if under the mechanism that favors high-effort agents the least, agents have the incentive to deviate to a lower cut-off, i.e., $\phi(s,\bs v_{\min}) \leq 0$, and under the  mechanism that favors high-effort agents the most, they have the incentive to deviate to a higher cut-off, i.e., $\phi(s,\bs v_{\max}) \geq 0$, there exists a mechanism $\bs v$ in between under which no agent has the incentive to deviate either upwards or downwards from $s$, i.e., $\phi(s,\bs v) = 0$. This means $s$ is a feasible equilibrium cut-off under some mechanism.

When $s=0$, only the downward deviation condition under $\bs v_{\min}$ is required, since the incentive to move to a lower cut-off also makes $s=0$ an equilibrium. Similarly, only the upward deviation condition under $\bs v_{\max}$ is required when $s=1$.

To obtain the feasible set of equilibrium cut-offs induce by mechanisms that satisfy the three assumptions, we let $\bs v_{\min}=\underline{\bs v}$ and $\bs v_{\max}=\bar{\bs v}$, which yields $\cl S = \cl S(\underline{\bs v},\bar{\bs v})$. \Cref{thm_feasible} is also applicable under constraints that do not necessarily take $\underline{\bs v}$ and $\bar{\bs v}$ as the upper and lower bounds. This flexibility is particularly helpful for further analysis of the model, especially when it is generalized to include additional constraints, such as uncertainty in performance.

\begin{corollary}
	$s\in \cl S$ if and only if $\phi(s,\bar{\bs v})\geq 0$ or $s=0$.
\end{corollary}

Always exerting low effort is a dominant strategy under the reversed contest, as high effort does not increase the probability of being selected and incurs a cost. This implies that agents will always have an incentive to deviate to a lower cut-off at any given cut-off level, which automatically satisfies the condition $\phi(s,\underline{\bs v}) \leq 0$ for all $s$. This fulfills the first requirement for a cut-off to be feasible.

\begin{corollary}
	$\cl S(\underline{\bs v}, \bs v_r) = \{0\}$, $\cl S(\bs v_r,\bar{\bs v}) = \cl S = \cl S(\underline{\bs v},\bar{\bs v})$.
\end{corollary}

Exerting low effort is already a dominant strategy under a completely random allocation. This can be verified by noting that $Q^r_i(H, \bs s) = Q^r_i(L, \bs s)$ in such an allocation. Consequently, we have
\begin{equation*}
	\phi(s,\bs v_r) = Q^r_i(H, \bs s) - c(s) - Q^r_i(L, \bs s) = -c(s) \leq 0,
\end{equation*}
where $Q^r_i$ denotes the interim allocation under the random mechanism. The inequality is strict when $s > 0$. Therefore, any mechanism with an allocation vector between $\underline{\bs v}$ and $\bs v_r$ is unhelpful to the principal, as it can only induce an equilibrium cut-off of $s = 0$. The principal can achieve the full feasible set $\cl S$ by only employing mechanisms with allocation vectors between $\bs v_r$ and $\bar{\bs v}$.

\begin{figure}[htbp]
	\centering
	\begin{tikzpicture}[yscale=0.9]
		\begin{axis}[
			axis lines=middle,
			ymin=-1.1, ymax=0.3,
			ytick distance=0.2,
			xlabel=$s$,
			ylabel={$\phi$},
			legend style={at={(1,.6)},anchor=north east}
		]
		
		\addplot [
			name path=g,
			domain=0:1,
			samples=100,
			color=blue,
		]{(x^4)/3-0.5*x+2/9};
		\addlegendentry{$\phi(s,\bar{\bs v})$}

		\addplot [
			name path=f,
			domain=0:1,
			samples=100,
			color=red,
			dashed,
			dash pattern=on 7.5pt off 2.5pt,
		]{(x^4)/3-0.5*x-7/9};
		\addlegendentry{$\phi(s,\underline{\bs v})$}

		\draw [black, line width=2pt] (0,0) -- (.47976448,0);
		\draw [black, line width=2pt] (.91809379,0) -- (1,0);
		\addlegendimage{color=black, line width=2pt}
		\addlegendentry{$\cl S(\underline{\bs v}, \bar{\bs v})$}

		\addplot [
			thick,
			color=blue,
			fill=white!95!yellow,
		]fill between[
			of=f and g,
			soft clip={domain=0:1},
		];

		\end{axis}
	\end{tikzpicture}
	\caption{An illustration of \Cref{thm_feasible}\\$n=3$, $m=2$, $F(x)=x^4$, $c(x)=\frac 12x+\frac 19$}
	\label{thm_feasible_il}
\end{figure}

\Cref{thm_feasible_il} provides an illustration for \Cref{thm_feasible}. As the allocation vector $\bs v$ increases from $\underline{\bs v}$ to $\bar{\bs v}$, the curve of $\phi(s,\bs v)$ transitions from the bottom to the top, passing through every point in the shaded area between them. The intersection of the horizontal axis with the shaded area represents the feasible set $\cl S(\underline{\bs v}, \bar{\bs v})$.

From this example, it's evident that the feasible set $\cl S$ is not necessarily a closed interval. Generally, the feasible set can consist of several disjoint closed intervals or even singletons. This unusual behavior arises entirely from \Cref{ass_bound}, which stipulates that each agent must receive no more than one unit of the prize. To understand this better, we start by examining the probability of being selected that an agent can secure in a standard contest by exerting high or low effort. These probabilities form the following ``per capita allocation vector,''
\begin{equation*}
	\begin{aligned}
		\left(\frac{\bar v_k}k\right)_{k=1}^{n-1}&=\left(\underbrace{1,1,\dots,1}_{m},\frac{m}{m+1},\frac{m}{m+2},\dots,\frac{m}{n-1}\right),\ \text{and}\\
		\left(\frac{m-\bar v_k}{n-k}\right)_{k=1}^{n-1}&=\left(\frac{m-1}{n-1},\frac{m-2}{n-2},\dots,\frac{1}{n-m+1},\underbrace{0,0,\dots,0}_{n-m}\right).
	\end{aligned}
\end{equation*}

When the cut-off $s$ is low such that the average number of agents exerting high effort, denoted by $k$, is less than $m$, the probability of these agents being selected remains at 1. Meanwhile, the probability of agents exerting low effort being selected decreases as $k$ increases. This creates an increasing incentive for agents to deviate towards a higher cut-off as $s$ increases, thereby driving up $k$. In other words, the term $\bar Q_i(H,\bs s) - \bar Q_i(L,\bs s)$ in $\phi(s,\bar{\bs v})$ tends to increase with $s$, where $\bar Q$ represents the interim allocation rule corresponding to $\bar{\bs v}$.

However, when the cut-off $s$ is high enough that $k$ is significantly larger than $m$, the probability of agents exerting low effort being selected drops to 0, while the probability of agents exerting high effort being selected decreases as $k$ increases. This happens because more agents are competing for a fixed amount of prizes. In this scenario, agents have an increasing incentive to deviate towards a lower cut-off, and the term $\bar Q_i(H,\bs s) - \bar Q_i(L,\bs s)$ in $\phi(s,\bar{\bs v})$ tends to decrease as $s$ rises. This non-monotonic behavior, combined with the structure of the cost function $c$, can cause $\phi(s,\bar{\bs v})$ to cross the horizontal axis (where $\phi=0$) multiple times, resulting in a discontinuous feasible set $\cl S$, as illustrated in \Cref{thm_feasible_il}.

In the specific case where there is only a single prize to allocate, any increase in the number of agents exerting high effort will lower the average probability of each being selected. Hence, any increase in the cut-off $s$ will only reduce the incentive for agents to deviate upwards. Here, $\phi(s,\bar{\bs v})$ tends to decrease monotonically with $s$, leading to a more regular structure of the feasible set $\cl S$. In fact, this scenario leads us to the following proposition.

\begin{proposition}
	\label{prop_interval}
	When $m=1$, the feasible set $\cl S$ is given by
	\begin{equation*}
		\cl S = \switchplus{
			\{0\}, & \phi(0,\bs 1){}&{}\leq 0\\
			{[}0,\phi^{-1}(0){]}, &\phi(1,\bs 1){}&{}<0<\phi(0,\bs 1)\\
			{[}0,1{]}, & \phi(1,\bs 1){}&{}\geq 0\\
		}{cr@{}l},
	\end{equation*}
	where $\bs 1=(1,1,\dots,1)\in[0,1]^{n-1}$ is equal to $\bar{\bs v}$ when $m=1$, and $\phi^{-1}$ is the inverse function of $\phi(s,\bs 1)$ with respect to $s$.
\end{proposition}

\begin{proof}
	See \Cref{app:prop_interval_pf}.
\end{proof}

\begin{save}[prop_interval_pf]
\begin{proof}
	When $m=1$, the allocation vector of the standard and the reversed contest are
	\begin{equation*}
		\begin{aligned}
			\underline{\bs v}&=\bs 0=(0,0,\dots,0),\text{ and}\\
			\bar{\bs v}&=\bs 1=(1,1,\dots,1).
		\end{aligned}
	\end{equation*}
	Plugging $\underline{\bs v}$ into \eqref{exp_phi}, we have 
	\begin{equation*}
		\phi(s,\bs 0) = -c(s)-\frac 1n\sum_{k=0}^{n-2}(F(s))^k \leq 0,
	\end{equation*}
	the inequality being strict for all $s>0$. By \Cref{thm_feasible}, $0\in\cl S$, and for any $s\in(0,1]$, $s\in\cl S$ is if and only if $\phi(s,\bs 1)\geq0$. Plugging $\bar{\bs v}$ into \eqref{exp_phi} yields
	\begingroup\allowdisplaybreaks
	\begin{align*}
		\phi(s,\bs 1) &= \frac 1n\sum_{k=1}^{n-1}(F(s))^{k-1}(1-F(s))^{n-1-k}\binom{n}{k}-c(s)-\frac 1n\sum_{k=0}^{n-2}(F(s))^k\\
		&= \frac{1-(F(s))^n-(1-F(s))^n}{nF(s)(1-F(s))} - c(s) - \frac {1-(F(s))^{n-1}}{n(1-F(s))} \\
		&= \frac{1-F(s)-(1-F(s))^n}{nF(s)(1-F(s))} - c(s)\\
		&= \frac{1-(1-F(s))^{n-1}}{nF(s)} - c(s)\\
		&= \frac 1n\sum_{k=0}^{n-2}(1-F(s))^k - c(s),
	\end{align*}
	\endgroup
	which is strictly decreasing in $s$, with potential jump discontinuities. Therefore, $\phi(s,\bar{\bs v})$ has a well defined inverse function $\phi^{-1}$ on $[\phi(1,\bs 1),\phi(0,\bs 1)] = [(n-1)/n-c(0),-c(1)]$ with respect to $s$. Whenever $\phi(s,\bar{\bs v})$ has a jump discontinuity from $y_1$ to $y_2$ ($y_1>y_2$) at $x$, define $\phi^{-1}(y)=x$ for all $y\in[y_2,y_1]$. $\phi^{-1}$ is also a decreasing function.

	When $\phi(0,\bs 1)\leq 0$, for all $s\in(0,1]$, 
	\begin{equation*}
		\phi(s,\bs 1)<\phi(0,\bs 1)\leq 0,
	\end{equation*}
	which implies none of $s\in(0,1]$ is feasible, i.e. $\cl S=\{0\}$. 
	
	When $\phi(1,\bs 1)<0<\phi(0,\bs 1)$, there exists $\phi^{-1}(0)\in(0,1)$ such that 
	\begin{equation*}
		\phi(s,\bs 1)\geq \phi(\phi^{-1}(0),\bs 1) = 0
	\end{equation*}
	for all $0\leq s\leq\phi^{-1}(0)$, and that
	\begin{equation*}
		\phi(s,\bs 1) < \phi(\phi^{-1}(0),\bs 1) = 0
	\end{equation*}
	for all $\phi^{-1}(0) < s \leq 1$. This implies $\cl S=[0,\phi^{-1}(0)]\subsetneq[0,1]$.

	When $\phi(1,\bs 1)\geq 0$, for all $s\in(0,1]$,
	\begin{equation*}
		\phi(s,\bs 1) \geq \phi(1,\bs 1)\geq 0,
	\end{equation*}
	which implies the whole interval $(0,1]$ is feasible, i.e. $\cl S=[0,1]$.

	Hence when $m=1$ (or when \Cref{ass_bound} is relaxed), the feasible set of equilibrium cut-offs is always a single interval.
\end{proof}
\end{save}

When there is only a single prize to allocate, or alternatively when \Cref{ass_bound} is relaxed, the feasible set $\cl S$ becomes a single interval that starts from $0$.

Although the feasible set $\cl S$ is always a subset of $[0,1]$, which is one-dimensional, the set of all allocation vectors has $n-1$ degrees of freedom. Intuitively, there are significantly more mechanisms available than equilibrium cut-offs. This insight underscores that the principal doesn't need to explore the full complexity of the mechanism space to achieve the desired outcomes. Instead, focusing on a carefully chosen subset of mechanisms suffices to span the full range of feasible equilibrium cut-offs.

\begin{definition}
	A {\it collection of mechanisms with one parameter} is a continuous mapping $v:[0,1]\to[0,m]^{n-1}$ from a single parameter $t$ to an allocation vector $v(t)$ that satisfies $v(0)=\underline{\bs v}$ or $\bs v_r$, and $v(1)=\bar{\bs v}$.
\end{definition}

\begin{example}[Blind-eye mechanisms]
	We introduce the ``blind-eye mechanisms,'' wherein the principal intentionally turns a blind eye to agents' high effort under certain conditions. The mechanism operates as follows:
	\begin{enumerate}
		\item The principal selects a probability $t\in[0,1]$, which is disclosed to the agents.
		\item Each agent decides whether to exert high or low effort. Low effort is always correctly observed as low effort. In contrast, high effort is correctly observed as high effort with probability $t$, and misobserved as low effort with probability $1-t$.
		\item The principal then conducts a standard contest based on the {\it observed} action profile.
	\end{enumerate}
	Blind-eye mechanisms are a collection of mechanisms with one parameter. When $t=0$, the mechanism reduces to a completely random allocation; when $t=1$, it is equivalent to a standard contest.
\end{example}

\begin{example}[Quota mechanisms]
	Suppose there is only a single prize. We define the ``quota mechanisms'', where the principal imposes a quota on the total probability that agents who exert high effort will be selected. The mechanism operates as follows:
	\begin{enumerate}
		\item The principal sets a quota $t\in[0,1]$, which is disclosed to the agents.
		\item Agents choose to exert either high or low effort. The prize is awarded to agents who exerted high effort as a group with probability $t$, and to those who exerted low effort as a group with probability $1-t$.
		\item The prize is subsequently allocated by a fair lottery within the prize-winning group.
	\end{enumerate}
	Quota mechanisms are a collection of mechanisms with one parameter that are applicable only when $m=1$ or when \Cref{ass_bound} is relaxed. When $t=0$, the mechanism functions as a reversed contest; when $t=1$, it aligns with a standard contest.
\end{example}

A collection of mechanisms with one parameter is a one-dimensional subset of all available mechanisms.

\begin{proposition}
	\label{prop_collection}
	Let $v:[0,1]\to[0,m]^{n-1}$ be a collection of mechanisms with one parameter. For any $s\in\cl S$, there exists $t\in[0,1]$ such that $s$ is an equilibrium cut-off induced by $v(t)$.
\end{proposition}

\begin{proof}
	See \Cref{app:prop_collection_pf}.
\end{proof}

\begin{save}[prop_collection_pf]
\begin{proof}
	\Cref{prop_collection} is a direct corollary from the proof of \Cref{thm_feasible}.

	By \Cref{thm_feasible} and the two corollaries, $0\in\cl S$, which is an equilibrium cut-off under $v(0)=\underline{\bs v}$ or $\bs v_r$. For all $s\in\cl S\backslash\{0\}$, 
	\begin{equation*}
		\phi(s,v(0))<0\leq \phi(s,\bar{\bs v})= \phi(s,v(1))
	\end{equation*}
	hold. Furthermore, $\phi(s,\bs v)$ is continuous in $\bs v$, and $v(t)$ is continuous in $t$ by assumption. They together imply that for any given cut-off $s\in\cl S\backslash\{0\}$, $\phi(s,v(t))$ is continuous in $t$. Hence by intermediate value theorem, there exists $t\in(0,1]$ such that $\phi(s,v(t))=0$, that is, $s$ is an equilibrium cut-off induced by allocation vector $v(t)$.
 \end{proof}
\end{save}

\Cref{prop_collection} establishes that the principal can achieve any feasible equilibrium cut-off and optimize any utility target using mechanisms from a specific one-dimensional subset of all available mechanisms. This proposition also highlights the significant flexibility in selecting such one-parameter collections of mechanisms. Unlike the optimal auction problem, where a particular set of mechanisms emerges as superior while others are suboptimal, our optimal selective contest problem does not exhibit inherently inferior mechanisms. Any allocation vector corresponds to some collection of mechanisms with one parameter, making it possible candidate for principal to implement the equilibrium cut-off that aligns with her preferences.

A noteworthy subtlety in this context is the potential for multiple equilibria to arise under certain mechanisms. In general, it is an intractable problem whether a mechanism that induces $s$ as its unique equilibrium cut-off always exists for any given $s\in\cl S$. However, the uniqueness issue becomes more straightforward when there is only a single prize to allocate or when \Cref{ass_bound} is relaxed.

\begin{proposition}
	\label{prop_quota}
	Let $v_Q:[0,1]\to[0,m]^{n-1}$ be the collection of quota mechanisms. If $c$ and $F$ are continuous on $[0,1]$, then for any $s\in\cl S$, there exists $t\in[0,1]$ such that $s$ is the unique equilibrium cut-off under $v_Q(t)$. When $0<s<1$, such $t$ is uniquely determined by $s$ as
	\begin{equation*}
		t(s)=\frac{\sum_{k=0}^{n-2}(F(s))^k+nc(s)}{\sum_{k=0}^{n-2}(F(s))^k+\sum_{k=0}^{n-2}(1-F(s))^k}.
	\end{equation*}
\end{proposition}

\begin{proof}
	See \Cref{app:prop_quota_pf}.
\end{proof}

\begin{save}[prop_quota_pf]
\begin{proof}
	For simplicity, we focus on the case where $m=1$. Cases where $m\geq 2$ and \Cref{ass_bound} is relaxed are equivalent to the single prize case with the effort cost scaled down to its $1/m$.

	\Cref{prop_exist} guarantees that there always exists $t$ such that $s$ is an equilibrium cut-off under $v_Q(t)$ for all $s\in\cl S$. We move on to show that for all $t\in[0,1]$, the equilibrium cut-off under $v_Q(t)$ is unique. We can write $\phi(s,v_Q(t))$ with respect to $s$ and $t$ explicitly as
	\begingroup\allowdisplaybreaks
	\begin{align*}
		\phi(s,v_Q(t)) &= \frac tn\sum_{k=1}^{n-1} (F(s))^{k-1}(1-F(s))^{n-1-k}\binom{n}{k} - c(s) - \frac 1n\sum_{k=0}^{n-2}(F(s))^k\\
		&= \frac{t(1-(F(s))^n-(1-F(s))^{n})}{nF(s)(1-F(s))} - c(s) - \frac{1-(F(s))^{n-1}}{n(1-F(s))} \\
		&= \frac{t(1-(1-F(s))^{n-1})}{nF(s)} - c(s) - \frac{(1-t)(1-(F(s))^{n-1})}{n(1-F(s))} \\
		&= \frac tn\sum_{k=0}^{n-2}(1-F(s))^k - \frac{1-t}n \sum_{k=0}^{n-2}(F(s))^k - c(s).
		\stepcounter{equation}
		\tag{\theequation}
		\label{exp_phi_vQ}
	\end{align*}
	\endgroup
	It is easy to verify that 
	\begin{equation*}
		\frac tn\sum_{k=0}^{n-2}(1-F(s))^k
	\end{equation*}
	and 
	\begin{equation*}
		-\frac{1-t}n\sum_{k=0}^{n-2}(F(s))^k
	\end{equation*}
	are both decreasing in $s$ and strictly increasing in $t$, and that $-c(s)$ is strictly decreasing in $s$. Hence, $\phi(s,v_Q(t))$ is strictly decreasing in $s$ and strictly increasing in $t$.

	For a given $t$, when $0$ is an equilibrium cut-off, $\phi(s,v_Q(t))<\phi(0,v_Q(t))\leq 0$, $\forall s\in(0,1]$, i.e. the only equilibrium cut-off is $0$; when $1$ is an equilibrium cut-off, $\phi(s,v_Q(t))>\phi(1,v_Q(t))\geq 0$, $\forall s\in[0,1)$, i.e. the only equilibrium cut-off is $1$; when some $s_0\in(0,1)$ is an equilibrium cut-off, $\phi(s_0,v_Q(t))=0$, which then implies by strict monotonicity that $\phi(s,v_Q(t))>0$ for all $s\in[0,s_0)$, and that $\phi(s,v_Q(t))<0$ for all $s\in(s_0,1]$, i.e. $s_0$ is the only equilibrium cut-off. Therefore, we proved the uniqueness of equilibrium cut-off under quota mechanisms.

	For a given $s\in(0,1)$, $s$ is an equilibrium cut-off under $v_Q(t)$ if and only if $\phi(s,v_Q(t))=0$. Solving this equation with the explicit form of $\phi$ given by \eqref{exp_phi_vQ} yields
	\begin{equation*}
		t = \frac{\sum_{k=0}^{n-2}(F(s))^k+nc(s)}{\sum_{k=0}^{n-2}(F(s))^k+\sum_{k=0}^{n-2}(1-F(s))^k}.
	\end{equation*}
\end{proof}
\end{save}

\Cref{prop_quota} offers a rationale for assessing the optimality of a mechanism based on the equilibrium it induces. It also outlines a method for determining an optimal contest design once the principal has identified the optimal equilibrium cut-off.

\section{Optimal Contest}
\label{sec:optimal}

For convenience, throughout this section, we assume that the type distribution $F$ has a positive density function $f$ on $[0,1]$, with at most countably many exceptions, and that the cost function $c$ is piecewise differentiable on $[0,1]$.

Given the feasible set of equilibrium cut-offs, the principal's optimization problem can be expressed as 
\begin{equation*}
	\max_{s}\ \eta-\lambda C,\quad\st s\in\cl S.
\end{equation*}

\Cref{thm_cutoff} implies that all feasible cost-efficiency pairs $(C,\eta)$ lie on a curve. Let the parametric equation of this curve be denoted by
\begin{equation*}
	\switch{
		C &= C(s) \\
		\eta &= \eta(s)
	}.
\end{equation*}
As the equilibrium cut-off increases, agents are more likely to exert high effort and, consequently, incur the associated effort cost. This can be verified through the explicit form of the societal cost given by \eqref{C_exp_cutoff},
\begin{equation*}
	C(s) = \int_0^s c(\theta)\D F(\theta), \tag{\ref{C_exp_cutoff}}
\end{equation*}
which shows that the societal cost $C$ is strictly increasing in the equilibrium cutoff. Therefore, there is a one-to-one correspondence between the equilibrium cutoff and societal cost, implying that $\eta$ is actually a function of $C$. We represent this relationship by
\begin{equation*}
	\eta = \Eta(C) = \eta(C^{-1}(C)),
\end{equation*}
where $C^{-1}$ is the inverse of $C(s)$. The principal's problem can then be reformulated as
\begin{equation}
	\label{target}
	\max_{C}\ \eta-\lambda C,\quad\st \eta=\Eta(C),\ C\in C(\cl S),
\end{equation}
where $C(\cl S)=\{C:C=C(s),\ s\in\cl S\}$ represents the set of all societal cost levels that can be achieved under some feasible equilibrium.

There are several advantageous properties that we can leverage from the explicit expression of the principal's problem. Our approach treats the problem as one of maximizing the intercept of a straight line that intersects a given curve, which may have an arbitrary shape, with certain portions removed due to constraints imposed by the feasible set of equilibrium cut-offs.

To address the problem, we define $\Eta_0$ as 
\begin{equation*}
	\Eta_0(C) = \Eta(C)\cdot\mathbbm{1}[C\in C(\cl S)].
\end{equation*}
Furthermore, let $\bar \Eta_0(C)$ be the concavification of $\bar\Eta_0$, defined as
\begin{equation*}
	\bar\Eta_0(C) = \max_{C_1,C_2}[\alpha\Eta_0(C_1) + (1-\alpha)\Eta_0(C_2)],
\end{equation*}
where $\alpha C_1+(1-\alpha)C_2=C$ and $0<\alpha<1$. Denote by $\bar\Eta_0^{\prime-1}$ the inverse of the derivative of $\bar\Eta_0$, i.e. $\bar\Eta_0'$. To extend the domain to $\mathbb{R}_+$, we introduce the following supplementary definitions:
\begin{enumerate}
	\item $\bar\Eta_0^{\prime-1}(y) := x$ if $x$ is a jump discontinuity of $\bar\Eta_0'$ from $y_1$ to $y_2$ and $y_2\leq y\leq y_1$.
	\item $\bar\Eta_0^{\prime-1}(y) := 0$ if $y\geq\bar\Eta_0'(0)$.
	\item $\bar\Eta_0^{\prime-1}(y) := x_1$ or $x_2$ if $\forall x\in(x_1,x_2)$, $\bar\Eta_0'(x)=y$ and $\forall x\notin[x_1,x_2]$, $\bar\Eta_0'(x)\neq y$.
\end{enumerate}

Our next main result provides a general solution to problem \eqref{target}.

\begin{theorem}
	\label{thm_optimal}
	An optimal societal cost level $C^*$ is given by
	\begin{equation*}
		C^* = \bar\Eta_0^{\prime-1}(\lambda).
	\end{equation*}
	A corresponding optimal equilibrium cut-off $s^*$ is given by
	\begin{equation*}
		s^* = C^{-1}(C^*).
	\end{equation*}
\end{theorem}

\begin{proof}
	See \Cref{app:thm_optimal_pf}.
\end{proof}

\begin{save}[thm_optimal_pf]
\begin{proof}
	We rewrite the optimization target as
	\begin{equation}
		\label{OPT_1}
		\max_{C}\ \Eta(C)-\lambda C,\quad\st C\in C(\cl S).
	\end{equation}
	We can merge the feasibility constraint into the target function as
	\begin{equation}
		\label{OPT_2}
		\max_{C}\ \Eta_0(C)-\lambda C.
	\end{equation}
	By \Cref{thm_feasible} and its corollaries, $0\in\cl S$, and therefore $0\in C(\cl S)$. For all $C\notin C(\cl S)$, $C>0$, 
	\begin{equation*}
		\Eta_0(C)-\lambda C = -\lambda C < 0 = \Eta_0(0)-\lambda\cdot 0.
	\end{equation*}
	For all $C\in C(\cl S)$,
	\begin{equation*}
		\Eta_0(C)-\lambda C = \Eta(C)-\lambda C.
	\end{equation*}
	Hence, problem \eqref{OPT_1} is equivalent to problem \eqref{OPT_2}.

	Consider the following problem:
	\begin{equation}
		\label{OPT_3}
		\max_{C}\ \bar\Eta_0(C)-\lambda C.
	\end{equation}
	We know that for all $C$, 
	\begin{equation*}
		\bar\Eta_0(C) = \max_{C_1,C_2}[\alpha\Eta_0(C_1) + (1-\alpha)\Eta_0(C_2)]\geq \alpha\Eta_0(C) + (1-\alpha)\Eta_0(C) = \Eta_0(C).
	\end{equation*}
	This gives
	\begin{equation}
		\label{COMP_1}
		\max_{C}[\bar\Eta_0(C)-\lambda C] \geq \max_{C}[\Eta_0(C)-\lambda C].
	\end{equation}
	\vspace*{0pt}

	\begin{claim*}
		There exists $C^*\in\argmax_{C}[\bar\Eta_0(C)-\lambda C]$ such that $\bar\Eta_0(C^*)=\Eta_0(C^*)$ for all $\lambda>0$.
	\end{claim*}
	\begin{proof}[Proof of the claim]
		\renewcommand{\qedsymbol}{}
		Suppose $C_0\in\argmax_{C}[\bar\Eta_0(C)-\lambda C]$. The claim is proved by taking $C^*$ to be $C_0$ if $\bar\Eta_0(C_0)=\Eta_0(C_0)$. When $\bar\Eta_0(C_0)>\Eta_0(C_0)$, by definition, there exist $C_1<C_2$ such that 
		\begin{equation*}
			\bar\Eta_0(C_0) = \alpha\Eta_0(C_1) + (1-\alpha)\Eta_0(C_2),\ \alpha C_1 + (1-\alpha)C_2 = C_0.
		\end{equation*} 
		We want to show that both $C_1$ and $C_2$ satisfy the conditions of such $C^*$.
	\end{proof}
	
	Suppose $\bar\Eta_0(C_1)>\Eta_0(C_1)$. Then, there exists $C_{11}<C_{12}$ such that 
	\begin{equation*}
		\bar\Eta_0(C_1) = \beta\Eta_0(C_{11}) + (1-\beta)\Eta_0(C_{12}),\ \beta C_{11} + (1-\beta)C_{12} = C_1.
	\end{equation*}
	This leads to a contradiction that 
	\begingroup\allowdisplaybreaks
	\begin{align*}
		\bar\Eta_0(C_0) &= \alpha\Eta_0(C_1) + (1-\alpha)\Eta_0(C_2)\\
		&<\alpha\bar\Eta_0(C_1) + (1-\alpha)\Eta_0(C_2)\\
		&=\alpha\beta\Eta_0(C_{11}) + \alpha(1-\beta)\Eta_0(C_{12}) + (1-\alpha)\Eta_0(C_2)\\
		&\leq\max_{\substack{
			C_1,C_2,C_3\\
			\gamma_1C_1+\gamma_2C_2+(1-\gamma_1-\gamma_2)C_3 = C_0\\
			0<\gamma_1,\gamma_2<1
		}}[\gamma_1\Eta_0(C_1)+\gamma_2\Eta_0(C_2)+(1-\gamma_1-\gamma_2)\Eta_0(C_3)]\\
		&=\max_{\substack{
			C_1,C_2\\
			\alpha C_1+(1-\alpha)C_2 = C_0\\
			0<\alpha<1
		}}[\alpha \Eta_0(C_1)+(1-\alpha)\Eta_0(C_2)]\\
		&= \bar\Eta_0(C_0).
	\end{align*}
	\endgroup
	Hence, $\bar\Eta_0(C_1)=\Eta_0(C_1)$. Similar arguments apply in showing $\bar\Eta_0(C_2)=\Eta_0(C_2)$.

	Furthermore, 
	\begingroup\allowdisplaybreaks
	\begin{align*}
		\max_{C}[\bar\Eta_0(C)-\lambda C]&=\bar\Eta_0(C_0)-\lambda C_0\\
		&= (\alpha\Eta_0(C_1) + (1-\alpha)\Eta_0(C_2)) - \lambda (\alpha C_1 + (1-\alpha)C_2)\\
		&= \alpha(\Eta_0(C_1) - \lambda C_1) + (1-\alpha)(\Eta_0(C_2) - \lambda C_2)\\
		&\leq \alpha\max_{C}[\bar\Eta_0(C)-\lambda C] + (1-\alpha)\max_{C}[\bar\Eta_0(C)-\lambda C]\\
		&= \max_{C}[\bar\Eta_0(C)-\lambda C],
	\end{align*}
	\endgroup
	which implies that
	\begin{equation*}
		\Eta_0(C_1) - \lambda C_1=\Eta_0(C_2) - \lambda C_2\max_{C}[\bar\Eta_0(C)-\lambda C],
	\end{equation*}
	i.e. $C_1,C_2\in\argmax_{C}[\bar\Eta_0(C)-\lambda C]$. 

	We have shown that both $C_1$ and $C_2$ satisfy the conditions of such $C^*$, which ultimately proves the claim. ({\it Proof of the claim ends.})
	\vspace{.75em}

	The claim implies
	\begin{equation*}
		\max_{C}[\bar\Eta_0(C)-\lambda C] \leq \max_{C}[\Eta_0(C)-\lambda C],
	\end{equation*}
	which together with \eqref{COMP_1} yields
	\begin{equation*}
		\max_{C}[\bar\Eta_0(C)-\lambda C] = \max_{C}[\Eta_0(C)-\lambda C]
	\end{equation*}
	for all $\lambda$, and they always have at least one common maximizer. 

	$\bar\Eta_0$ is concave and therefore piecewise differentiable. The first order condition of problem \eqref{OPT_3} is
	\begin{equation*}
		\bar\Eta_0'(C) - \lambda = 0,
	\end{equation*}
	where $\bar\Eta_0'$ is monotone decreasing. Let $\cl C_\lambda$ be the set $\{C:\bar\Eta_0'(C)=\lambda\}$. It can be either \begin{enumerate*}[label=(\roman*)]
		\item a singleton,
		\item an empty set, or
		\item an interval.
	\end{enumerate*}

	When $\cl C_\lambda=\{\bar\Eta_0^{\prime-1}(\lambda)\}$ is a singleton, let $C^* = \bar\Eta_0^{\prime-1}(\lambda)$, which uniquely maximizes \eqref{OPT_3}. According to the claim, it is also an optimizer to problem \eqref{OPT_2} and therefore problem \eqref{OPT_1}.

	When $\cl C_\lambda=\varnothing$, two cases may occur. If there exists a jump discontinuity of $\bar\Eta_0'$ from $\lambda_1$ to $\lambda_2$, denoted by $C_0$, such that $\lambda_2
	\leq\lambda\leq\lambda_1$, then $\bar\Eta_0'(C) - \lambda > 0$ for all $C<C_0$, and $\bar\Eta_0'(C) - \lambda < 0$ for all $C>C_0$. This means that $C_0$ is the unique maximizer to problem \eqref{OPT_3}. Problem \eqref{OPT_2} and \eqref{OPT_1} are solved by taking $C^*$ to be $C_0$. If instead $\lambda\geq\bar\Eta_0'(0)$, then $\bar\Eta_0'(C) - \lambda < 0$ for all $C>0$. This implies that $0$ is the unique maximizer to problem \eqref{OPT_3}, and our problems can be solved by letting $C^*=0$.

	When $\cl C_\lambda$ is an interval from $C_1$ to $C_2$, every point in $\cl C_\lambda$ solves problem \eqref{OPT_3}.\footnote{We do not explicitly write out the interval because either end of it could be open or closed.} According to the proof of the claim, we know for sure that the two endpoints $C_1$ and $C_2$ satisfy $\bar\Eta_0(C_1)=\Eta_0(C_1)$ and $\bar\Eta_0(C_2)=\Eta_0(C_2)$. Therefore both of them are maximizers of problem \eqref{OPT_2}. Our problems are solved by taking $C^*$ to be either $C_1$ or $C_2$.

	An optimal equilibrium cut-off $s^*$ satisfies $C(s^*)=C^*$, which gives
	\begin{equation*}
		s^* = C^{-1}(C^*).
	\end{equation*}
\end{proof}
\end{save}

An intuitive way to understand \Cref{thm_optimal} is by visualizing the principal's target function as a straight line ``dropping'' from above towards the curve $\eta=\Eta(C)$, or more precisely, towards the point set $\cl P=\{(C,\eta):\eta=\Eta(C),\ C\in C(\cl S)\}$. This process can be described as gradually lowering the principal's payoff $u$ from $+\infty$ to $0$, stopping at the exact moment when the line $\eta=\lambda C+u$ first touches $\cl P$. This line divides the plane into two half-planes, one of which entirely contains $\cl P$. The corresponding value of $u$ represents the maximum payoff the principal can achieve. 

As $\lambda$ varies, the common intersection of all such half-planes that contain $\cl P$ forms the convex hull of $\cl P$, whose upper boundary is precisely $\bar{\Eta}_0$, the concavification of $\Eta_0$, as defined in \Cref{thm_optimal}. The process of ``dropping'' the line reveals that the principal does not attain the optimal point within the interior of the convex hull of $\cl P$. Therefore, the problem can be equivalently expressed as
\begin{equation*}
	\max_{C}\ \eta-\lambda C,\quad\st \eta=\bar\Eta_0(C).
\end{equation*}
This problem is relatively straightforward to solve since $\bar\Eta_0$ is concave, allowing the maximizer to be determined via the first-order condition.

\begin{figure}[htbp]
	\centering
	\begin{subfigure}[b]{0.48\textwidth}
		\centering
		\begin{tikzpicture}
			\begin{axis}[
				axis lines=middle,
				xlabel=$C$,
				ylabel={$\eta$},
				legend style={at={(.83,.5)},anchor=north east},
				ticks=none,
			]
	
			\addplot [
				domain=0:1,
				samples=100,
				color=red,
				dashed,
				line width=1pt,
			]{16*x^5 - 55*x^4 + 63*x^3 - 30*x^2 + 6*x};
			\addlegendentry{$\Eta\ (\Eta\neq\Eta_0)$}
	
			\addplot [
				domain=0:0.05,
				samples=100,
				color=red,
				line width=1pt,
			]{16*x^5 - 55*x^4 + 63*x^3 - 30*x^2 + 6*x};
			\addlegendentry{$\Eta_0$}
			
			\addplot [
				domain=0.05:0.15,
				samples=100,
				color=blue,
				line width=1pt,
			]{1.7846*x + 0.14330635};
			\addlegendentry{$\bar\Eta_0\ (\bar\Eta_0\neq\Eta_0)$}
			
			\addplot [
				domain=0:0.05,
				samples=100,
				color=black,
				line width=5pt,
			]{0};
			\addlegendentry{$C(\cl S)$}
			
			\addplot [
				domain=0.15:1,
				samples=100,
				color=red,
				line width=1pt,
			]{16*x^5 - 55*x^4 + 63*x^3 - 30*x^2 + 6*x};
			\addplot [
				domain=0.05:0.15,
				samples=100,
				color=red,
				line width = 2pt,
			]{0};
			\addplot [
				domain=0.15:1,
				samples=100,
				color=black,
				line width=5pt,
			]{0};
			\addplot[color=black, dashed, line width=1pt] table[row sep = crcr]{0.05 0 \\ 0.05 0.23253625 \\};
			\addplot[color=black, dashed, line width=1pt] table[row sep = crcr]{0.15 0 \\ 0.15 0.41099625 \\};
			\addplot [
				domain=0.164229:0.703326,
				samples=100,
				color=blue,
				line width=1pt,
			]{0.32752*x + 0.3634};
			\end{axis}
		\end{tikzpicture}
		\caption{}
		\label{thm_optimal_il_left}
	\end{subfigure}
	\hfill
	\begin{subfigure}[b]{0.48\textwidth}
		\centering
		\begin{tikzpicture}
			\begin{axis}[
				axis lines=middle,
				xlabel=$C$,
				ylabel={$\eta$},
				legend pos=north east,
				ticks=none,
				legend columns=2, 
			]
	
			\addplot [
				domain=0:0.05,
				samples=100,
				color=red,
				line width=1pt,
			]{80*x^4 - 220*x^3 + 189*x^2 - 60*x + 6};
			\addlegendentry{}
	
			\addplot [
				domain=0.05:0.15,
				samples=100,
				color=blue,
				line width=1pt,
			]{1.7846};
			\addlegendentry{$\bar\Eta_0'$}
			
			\addplot [
				domain=0.15:0.164229,
				samples=100,
				color=red,
				line width=1pt,
			]{80*x^4 - 220*x^3 + 189*x^2 - 60*x + 6};
			\addlegendentry{}
	
			\addplot[color=red, dashed, line width=1pt] table[row sep = crcr]{0.05 1.7846 \\ 0.05 3.4455 \\};
			\addlegendentry{$\bar\Eta_0^{\prime-1}$}
	
			\addplot [
				domain=0.164229:0.703326,
				samples=100,
				color=blue,
				line width=1pt,
			]{0.32752};
	
			\addplot [
				domain=0.703326:1,
				samples=100,
				color=red,
				line width=1pt,
			]{80*x^4 - 220*x^3 + 189*x^2 - 60*x + 6};
			
			\addplot[color=red, dashed, line width=1pt] table[row sep = crcr]{0.15 1.7846 \\ 0.15 0.5505 \\};

			\end{axis}
		\end{tikzpicture}		
		\caption{}
		\label{thm_optimal_il_right}
	\end{subfigure}
	\caption{An illustration of \Cref{thm_optimal}}
	\label{thm_optimal_il}
\end{figure}

\Cref{thm_optimal_il} provides an illustration for \Cref{thm_optimal}. As shown in \Cref{thm_optimal_il_left}, we first eliminate the infeasible points by lowering them to $0$, resulting in $\Eta_0$. Next, we use straight lines to cover each segment of $\Eta_0$ that either drops to $0$ or lacks concavity, yielding the concavification $\bar\Eta_0$, which corresponds to the convex hull of the set $\{(C,\eta):\eta\leq\Eta_0(C)\}$. \Cref{thm_optimal_il_right} provides a sketch of the derivative of $\bar\Eta_0$. The horizontal flat regions correspond to areas where $\bar\Eta_0\neq\Eta_0$. These regions become the range of discontinuities, in which the inverse derivative $\bar\Eta_0^{\prime-1}$ does not take any value. This ensures that $\bar\Eta_0^{\prime-1}(\lambda)$ is always attainable in the original problem. 

Since $\bar{\Eta}_0^{\prime-1}(0)$ is always finite, when the principal assigns a sufficiently high weight to societal costs, i.e., when $\lambda$ is sufficiently large, it is optimal for the principal to incorporate randomness into the selection mechanism rather than employing the standard winner-takes-all contest, $\bar{\boldsymbol{v}}$.

To further explore the trade-off between societal cost and selection efficiency, we derive the explicit form of their budget constraint.

Recall that $C$ and $\eta$ are determined by equilibrium cut-off $s$ as
\begin{align*}
	C &= \int_0^s c(\theta)\D F(\theta),\text{ and}\\
	\eta &= \frac {nc(s)}{m\mu}\int_0^s(\mu-\theta)\D F(\theta).
\end{align*} 

This gives
\begingroup\allowdisplaybreaks
\begin{align*}
	\Eta' = \frac{\D \eta}{\D C} = \frac{\D \eta/\D s}{\D C/\D s}&=\frac{\frac{n}{m\mu}c(s)(\mu-s)f(s)+c'(s)\int_0^s(\mu-\theta)\D F(\theta)}{c(s)f(s)}\\
	&= \frac{n}{m\mu}\left[(\mu-s) + \frac{c'(s)}{c(s)f(s)}\int_0^s(\mu-\theta)\D F(\theta)\right] \\
	&= \frac{n}{m\mu} [(\mu-s) + \epsilon(s)\cdot(\mu-\E_{\theta\sim F}[\theta|\theta<s])], 
	\label{budget}
	\stepcounter{equation}
	\tag{\theequation}
\end{align*}
\endgroup
where
\begin{equation*}
	\epsilon(s) = \frac{c'(s)F(s)}{c(s)f(s)} = \left.\frac{\D\ln c}{\D\ln F}\right|_s
\end{equation*}
is the percentage change in effort cost with respect to a change in agents' type percentile.

\eqref{budget} characterizes the budget line between societal cost and selection efficiency. A small increase in societal cost has two effects on selection efficiency. First, agents with the marginal type $s$ who originally exert low effort now exert high effort, making them more likely to be selected. This improves selection efficiency if and only if the marginal agent's type $s$ is below the average type $\mu$---that is, if the marginal agent's ability is above average. This effect is captured by the $(\mu - s)$ term, which can be either positive or negative.

Second, to incentivize a higher equilibrium cutoff, the principal must increase the probability of selecting agents who exert high effort, which directly enhances selection efficiency. This effect is captured by the $\epsilon(s)\cdot(\mu-\E_{\theta\sim F}[\theta|\theta<s])$ term, which is always non-negative. An increase in societal cost will benefit selection efficiency only if the second effect aligns with or outweighs the first.

In the special case where $\Eta$ is concave, $\bar\Eta_0'(C)$ coincides with $\Eta'(C)$ for all $C\in C(\cl S)$. According to \Cref{thm_optimal}, the optimal equilibrium cut-off in this case is given by
\begin{equation*}
	s^* = \switch{
		&0, &1+\epsilon(0) &< \frac mn\lambda\\
		&s_0, &1+\epsilon(0) &\geq \frac mn\lambda\text{ and }s_0\in\cl S\\
		&\argmax_{s\in\{s_1,s_2\}}[\eta(s)-\lambda C(s)], &1+\epsilon(0) &\geq \frac mn\lambda\text{ and }s_0\notin\cl S\\
	},
\end{equation*}
where $s_0$ satisfies $\Eta'(C(s_0))=\lambda$, $s_1 = \max \cl S\cap[0,s_0]$, and $s_2 = \min \cl S\cap[s_0,1]$.

\section{Comparative Statics}
\label{sec:comparative}

In this section, we assume that $c$ and $F$ take the following form:

\begin{equation*}
	\begin{aligned}
		F(x) &= x^\alpha,\\
		c(x) &= \gamma(F(x))^\epsilon = \gamma x^{\alpha\epsilon},
	\end{aligned}
\end{equation*}
where $\alpha,\gamma,\epsilon>0$.

Given this function form, we have
\begin{equation*}
	\mu = \frac{\alpha}{1+\alpha},\ \epsilon(s)\equiv\epsilon,
\end{equation*}
which yields
\begin{equation*}
	\begin{aligned}
		\eta &= \frac nm\cdot\gamma s^{(\epsilon+1)\alpha}(1-s),\\
		C &= \frac{\gamma}{\epsilon+1}\cdot s^{(\epsilon+1)\alpha},\\
		\Eta'(C(s)) &= \frac n{m\alpha}((\epsilon+1)\alpha-((\epsilon+1)\alpha+1)s).
	\end{aligned}
\end{equation*}

$\Eta'$ is linear in $s$ and decreasing, indicating that $\Eta$ is concave. Therefore $\Eta=\bar\Eta_0$ wherever $C\in C(\cl S)$. We begin by considering the relaxed problem, without imposing the feasibility constraint on the equilibrium cut-off. The optimal equilibrium cut-off under this relaxed problem is $s^\star$ such that:
\begin{equation*}
	\Eta'(C(s^\star)) = \lambda,
\end{equation*}
which results in
\begin{equation}
	\label{sstar}
	s^\star=\frac{\left(\epsilon+1-\dfrac mn \lambda\right)\alpha}{(\epsilon+1)\alpha+1}.
\end{equation}

$s^\star$ is decreasing in $\lambda$, which reflects how much the principal values societal cost relative to selection efficiency. As the principal's concern for societal cost increases, she becomes less inclined to encourage agents to exert high effort. Consequently, the principal will employ a mechanism with a lower equilibrium cutoff when $\lambda$ increases.

$s^\star$ also decreases with $m/n$, the average number of prizes per agent. The intuition here is that as the number of selected agents increases, the likelihood that the principal will have to select agents with relatively low ability also increases, especially since the principal cannot withhold prizes. As a result, selection efficiency decreases given the same level of societal cost. To mitigate this inefficiency, the principal opts for a lower equilibrium cutoff, which reduces societal cost.

$s^\star$ is non-decreasing in $\alpha$ if and only if
\begin{equation}
	\label{s_alpha}
	\epsilon + 1 \geq \frac mn\lambda.
\end{equation}
To prove this, we take the derivative of $s^\star$ with respect to $\alpha$:
\begin{equation*}
	\frac{\partial s^\star}{\partial\alpha} = \frac{\epsilon + 1-\dfrac mn\lambda}{((\epsilon+1)\alpha+1)^2}.
\end{equation*}
This derivative is non-negative as long as \eqref{s_alpha} holds. Intuitively, $\alpha$ represents the skewness of the distribution of agents' types. A larger $\alpha$ indicates a higher density of agents with higher effort costs. This has two effects on the principal's payoff. On one hand, with the same equilibrium cut-off, an increase in $\alpha$ reduces the expected number of agents exerting high effort, thereby lowering societal cost. This reduction in societal cost motivates the principal to shift to a higher equilibrium cut-off. On the other hand, the increased density of lower-ability agents decreases selection efficiency for a given equilibrium cut-off, prompting the principal to adopt a lower equilibrium cut-off to ``sharpen'' the signal. When \eqref{s_alpha} holds, the first effect is more pronounced, leading the principal to increase the equilibrium cut-off as $\alpha$ rises. Otherwise, the principal is likely to decrease the equilibrium cut-off as $\alpha$ increases.

It is easy to verify that $s^\star$ is increasing in $\epsilon$ given \eqref{sstar}. The intuition is that a larger $\epsilon$ indicates a greater percentage change in agents' effort costs relative to their type percentile. This larger change enhances the ability to distinguish between agents of high and low ability, thereby increasing the marginal benefit of raising the equilibrium cut-off for selection efficiency. Consequently, with a higher $\epsilon$, the principal opts for a larger equilibrium cut-off.

$s^\star$ remains constant as $\gamma$ varies, indicating that the principal is concerned only with the rate of increase in agents' effort costs, rather than its absolute magnitude.

It is generally challenging to specify how the feasible set of equilibrium cut-offs changes in response to the parameters, especially given the potential complexity of its shape. To simplify the analysis, we focus on the case where there is only a single prize. According to \Cref{prop_interval}, the feasible set of equilibrium cutoffs is now a single interval $\cl S = [0,s_{\max}]$, where $s_{\max} = \max\{0, \min\{1, \phi^{-1}(0)\}\}$. When $s_{\max} = \phi^{-1}(0)$, we have
\begin{equation*}
	\begin{aligned}
		&\phi(s_{\max},m,m,\dots,m) = 0\\
		\iff& c(s_{\max})-\frac mn \sum_{k=0}^{n-2}(1-F(s))^k = 0\\
		\iff& \gamma (F(s))^{\epsilon}-\frac mn \sum_{k=0}^{n-2}(1-F(s))^k = 0\\
		\iff & \gamma s_{\max}^{\alpha\epsilon}-\frac mn \sum_{k=0}^{n-2}(1-s_{\max}^\alpha)^k = 0.
	\end{aligned}
\end{equation*}
The optimal equilibrium cut-off $s^*$ is then given by
\begin{equation*}
	s^* = \argmin_{s\in\cl S}|s-s^\star| = \switch{
		&0, & s^\star &< 0\\
		&s^\star, & s^\star&\in\cl S\\
		&s_{\max}, & s^\star&> s_{\max}
	}.
\end{equation*}

It is easy to verify that $s_{\max}$ is
\begin{enumerate*}[label=(\roman*)]
	\item increasing in $m/n$,
	\item increasing in $\alpha$,
	\item decreasing in $\gamma$, and
	\item increasing in $\epsilon$.
\end{enumerate*}
If we instead look at how $F(s_{\max})$, the percentile of the maximum equilibrium cut-off, changes according to these parameters, we find that $F(s_{\max})$ is
\begin{enumerate*}[label=(\roman*)]
	\item increasing in $m/n$,
	\item constant when $\alpha$ changes, 
	\item decreasing in $\gamma$, and
	\item increasing in $\epsilon$.
\end{enumerate*}

Intuitively, as $m/n$ increases, the average probability of being selected rises, which motivates more agents to exert high effort in the standard contest, leading to a higher $s_{\max}$. Conversely, when $\gamma$ increases, the effort cost for each agent rises, reducing the number of motivated agents and resulting in a lower $s_{\max}$. An increase in $\epsilon$ has the opposite effect to $\gamma$ because it lowers the cost of exerting high effort for every agent, thereby increasing $s_{\max}$.

Effort cost $c$ is assumed to be a power function of $F$, the percentile of the agents' types, and is not directly affected by $\alpha$ except through its influence on $F$. Since the analysis of $s_{\max}$ occurs at the equilibrium level, which is unaffected by the principal's payoff, we can relabel the agents' types $\theta_i$ as $F(\theta_i)$ and obtain the same equilibrium result. This implies that the percentile of $s_{\max}$, $F(s_{\max})$, remains unchanged under different forms of $F$.

Changes in $s_{\max}$ can influence the final optimal cutoff $s^*$ by providing the principal with more or less flexibility in selecting equilibrium cutoffs, even when the optimal solution to the unconstrained problem, $s^\star$, remains unchanged. While $\gamma$ does not affect $s^\star$, an increase in $\gamma$ can lead to a decrease in $s^*$ due to the contraction of the feasible set $\cl S$. The effect of the average chance of being selected, $m/n$, on $s^*$ is more complex. In some cases, $s^*$ may initially increase and then decrease as $m/n$ increases. The initial increase is due to a higher $s_{\max}$ when $s^\star$ is still greater than $s_{\max}$, while the subsequent decrease occurs when $s^\star$ continues to decrease and falls below $s_{\max}$.

\section{Conclusion}
\label{sec:conclusion}

This paper explores how to balance selection efficiency and agents' welfare, or societal cost, through contest design. We find that the expected selection efficiency and societal cost of a mechanism are fully reflected in the induced equilibrium strategy, independent of the mechanism's specific implementation and allocation outcome. We characterize the feasible set of possible equilibrium strategies and develop a simple ironing method to optimize the principal's payoff over this feasible set. Additionally, we examine the conditions under which a collection of mechanisms is sufficient as a policy toolbox, enabling a principal with any preference to find an optimal contest design from the collection.

Our main finding is that selection efficiency and societal cost are almost always complementary. Improving selection efficiency typically requires inducing more effort costs from contestants, and vice versa. This suggests that most mechanisms are neither dominant nor dominated. A significant proportion of mechanisms can be optimal, as long as they align with the principal's preferences. Another key implication is that to balance selection efficiency and societal cost, the principal must intentionally introduce randomness into the contest. The right amount of randomness can prevent excessive competition without significantly compromising efficiency.

Nevertheless, when the contest designer values contestants' welfare, it is generally suboptimal to base selection solely on a one-dimensional performance ranking. In real-world selective competitions, such as school admissions, policymakers might benefit from adopting a multidimensional evaluation system that provide opportunities to candidates who may appear less able, particularly when candidates are already well-prepared, and their additional effort is unlikely to significantly enhance their abilities in the long run.

\clearpage
\printbibliography

\clearpage 
\begin{appendices}

\crefalias{section}{appendix}
\crefalias{subsection}{appendix}
\renewcommand{\thesubsection}{\Alph{section}.\arabic{subsection}} 
\renewcommand{\theequation}{\Alph{section}.\arabic{equation}}
\newcommand{\topref}[2]{\texorpdfstring{\Cref{#2}}{#1{}\ref{#2}}}
\section{Proofs}

\subsection{Proof of \topref{Proposition}{prop_exist}}
\label{app:prop_exist}
\load{prop_exist_pf}

\subsection{Proof of \topref{Theorem}{thm_cutoff}}
\label{app:thm_cutoff}
\load{thm_cutoff_pf}

\subsection{Proof of \topref{Theorem}{thm_feasible}}
\label{app:thm_feasible_pf}
\load{thm_feasible_pf}

\subsection{Proof of \topref{Proposition}{prop_interval}}
\label{app:prop_interval_pf}
\load{prop_interval_pf}

\subsection{Proof of \topref{Proposition}{prop_collection}}
\label{app:prop_collection_pf}
\load{prop_collection_pf}

\subsection{Proof of \topref{Proposition}{prop_quota}}
\label{app:prop_quota_pf}
\load{prop_quota_pf}

\subsection{Proof of \topref{Theorem}{thm_optimal}}
\label{app:thm_optimal_pf}
\load{thm_optimal_pf}

\end{appendices}

\end{document}